# Controlled Oxidation of Tellurium Nanoparticles for Telecom Detection

Biman Jana*

Indian Institute of Science, Bangalore – 560012, India

*Corresponding Author, Email: biman.jana1050@gmail.com

ABSTRACT:

Surface oxidation plays an important role in governing charge transport in semiconductor nanoparticle assemblies. Here, we investigate the optoelectronic properties of solution-processed tellurium (Te) nanoparticle ensembles prepared through gradual ambient-air exposure. Structural and spectroscopic analyses reveal interconnected elemental Te nanoparticles with coexisting α-$TeO_2$ and hydroxyl-containing surface species, exhibiting a lower relative contribution of oxidized Te than previously investigated forced-air-oxidized ensembles. The resulting photodetectors demonstrate photoresponse at 405, 532, and 1550 nm, with maximum responsivities of 10.1, 14.4, and 8.5 mA/W, respectively. At 1550 nm, the device exhibits rise and fall times of 17.7 and 40.2 s, respectively, and an estimated shot-noise-limited specific detectivity of $1.3 \times 10^6$ Jones. These findings highlight the importance of controlling surface oxidation to balance interparticle charge transport and dark-current suppression, offering insights for the development of solution-processed Te-based photodetectors for short-wave infrared applications.

INTRODUCTION:

Surface-defect passivation is an effective strategy for improving the optoelectronic properties of semiconductors.[1–4] In one-dimensional (1D) semiconductor nanostructures, surface oxidation has been explored as a means of modifying surface states and suppressing defect-mediated carrier recombination.[5–8] However, the benefits of oxide passivation may not translate directly to semiconductor assemblies comprising discrete nanoparticles. Unlike continuous thin films or bulk crystals, nanoparticle ensembles contain numerous interparticle interfaces that photogenerated charge carriers must traverse before reaching the electrodes. Consequently, high optical absorption alone is insufficient to ensure efficient photodetection. The passivation layer must suppress surface-related losses without introducing substantial interfacial barriers that impede charge transport between neighboring nanoparticles.[9] Balancing surface-defect passivation with interparticle electrical connectivity is therefore essential for optimizing the optoelectronic performance of nanoparticle-based devices.

Tellurium (Te) is an intrinsically p-type elemental semiconductor that has attracted considerable interest in optoelectronic applications because of its narrow bandgap, high hole mobility, anisotropic crystal structure, and broad spectral response extending from the visible to the infrared. Its distinctive electronic and optical properties arise, in part, from its helical atomic chains aligned along the crystallographic c-axis, which are coupled through relatively weak van der Waals interactions. These structural characteristics make low-dimensional Te nanostructures attractive for photodetection and other optoelectronic applications. Accordingly, Te nanowires, nanorods, nanosheets, nanoparticles, and nanofilms have been investigated for photodetection across a broad spectral range.[10–12] Nevertheless, translating the intrinsic optoelectronic properties of individual Te nanostructures into functional nanoparticle-based devices presents additional challenges because charge carriers must traverse multiple interparticle junctions. In such assemblies, charge transport depends not only on the properties of the individual nanostructures but also on their surface chemistry, interparticle coupling, and electrical connectivity.[13–16]

Surface oxidation offers a potential route to passivating defects in Te nanostructures, but its extent must be carefully controlled to preserve charge transport across the resulting nanoparticle network. In our previous work, oxide-passivated Te nanoparticles enabled ensemble-level photodetection, demonstrating that surface oxidation can be compatible with photoresponse in nanoparticle assemblies.[8,17] However, although oxide formation can passivate surface defects, excessive oxidation may introduce additional barriers at interparticle contacts and thereby limit charge collection[18] . This raises an important materials-design question: how can the oxidation of Te nanostructures be controlled to retain the benefits of surface passivation while maintaining sufficiently connected pathways for photogenerated carriers? Oxide formation kinetics can determine the number of defects present in the system and can affect the charge carrier mobility across the oxide layer.[19–21]

Here, we adopt a slow oxidation process and improve the responsivity of Te nanoparticle-based photoconduction. Rather than forcefully injecting laboratory air containing oxygen into the precursor solution, as in our previous procedure,[8] we allow ambient laboratory air, containing approximately 20.9% oxygen by volume, to enter the reaction vessel without forced injection. The reaction is extended to seven days, providing a different oxidation environment from the more rapid oxidation employed previously. We investigate how this slow exposure to oxygen influences the resulting nanostructures and their photodetection characteristics. The resulting devices exhibit higher responsivity and external quantum efficiency at telecom wavelength, although their detectivity decreases because of increased dark current. These results highlight the importance of considering surface passivation, interparticle charge transport, and dark-current contributions together when optimizing oxide-passivated Te nanoparticle ensembles for photodetection.

RESULT AND DISCUSSION

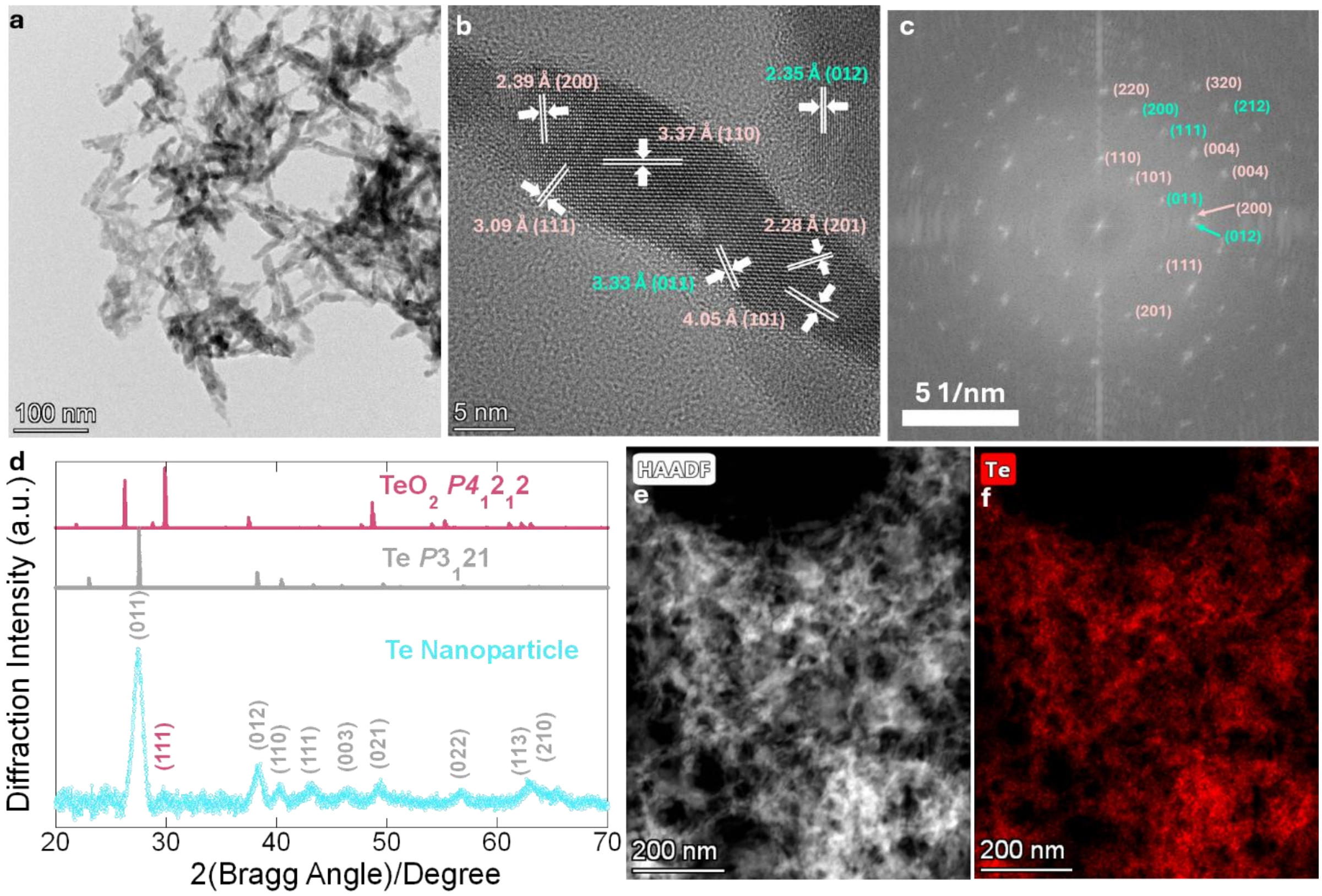

**Figure 1.** Structural and compositional characterization of tellurium nanoparticles (TNPs). **(a)** Bright-field transmission electron microscopy (BF-TEM) image of the TNPs. **(b)** High-resolution TEM (HRTEM) image of a representative TNP showing lattice fringes corresponding to Te and α- $TeO_2$. **(c)** Fast Fourier transform (FFT) pattern derived from the HRTEM image in (b). The Miller indices labelled in magenta and green correspond to α- $TeO_2$ and Te, respectively. **(d)** X-ray diffraction (XRD) pattern of a film prepared from the TNPs, showing reflections associated with Te and α-$TeO_2$. **(e)** High-angle annular dark-field scanning transmission electron microscopy (HAADF-STEM) image of the TNPs. **(f)** Corresponding energy-dispersive X-ray spectroscopy (EDS) elemental map showing the spatial distribution of Te.

The interconnected tellurium nanoparticles (TNPs) are observed in the bright-field transmission electron microscopy (BF-TEM) image (Figure 1a and Figure S1). Figure 1b presents a high-resolution TEM (HRTEM) image of a representative TNP, in which lattice fringes corresponding to Te and α-$TeO_2$ are identified. The fast Fourier transform (FFT) pattern derived from the same image (Figure 1c) exhibits diffraction spots consistent with the observed lattice fringes. These features are also consistent with the selected-area electron diffraction (SAED) pattern of the TNP ensemble (Figure S2). The X-ray diffraction (XRD) pattern, as shown in Figure 1d, further supports the presence of crystalline Te and α-$TeO_2$ through reflections corresponding to their respective crystallographic planes. However, the oxide-associated diffraction peaks are considerably weaker than those of Te, consistent with a comparatively small contribution from the crystalline oxide phase. The high-angle annular dark-field scanning transmission electron microscopy (HAADF-STEM) image (Figure 1e), together with the corresponding energy-dispersive X-ray spectroscopy (EDS) elemental maps, confirms the presence and spatial distribution of Te in the observed nanoparticles. Additional EDS elemental maps are provided in Figure S3.

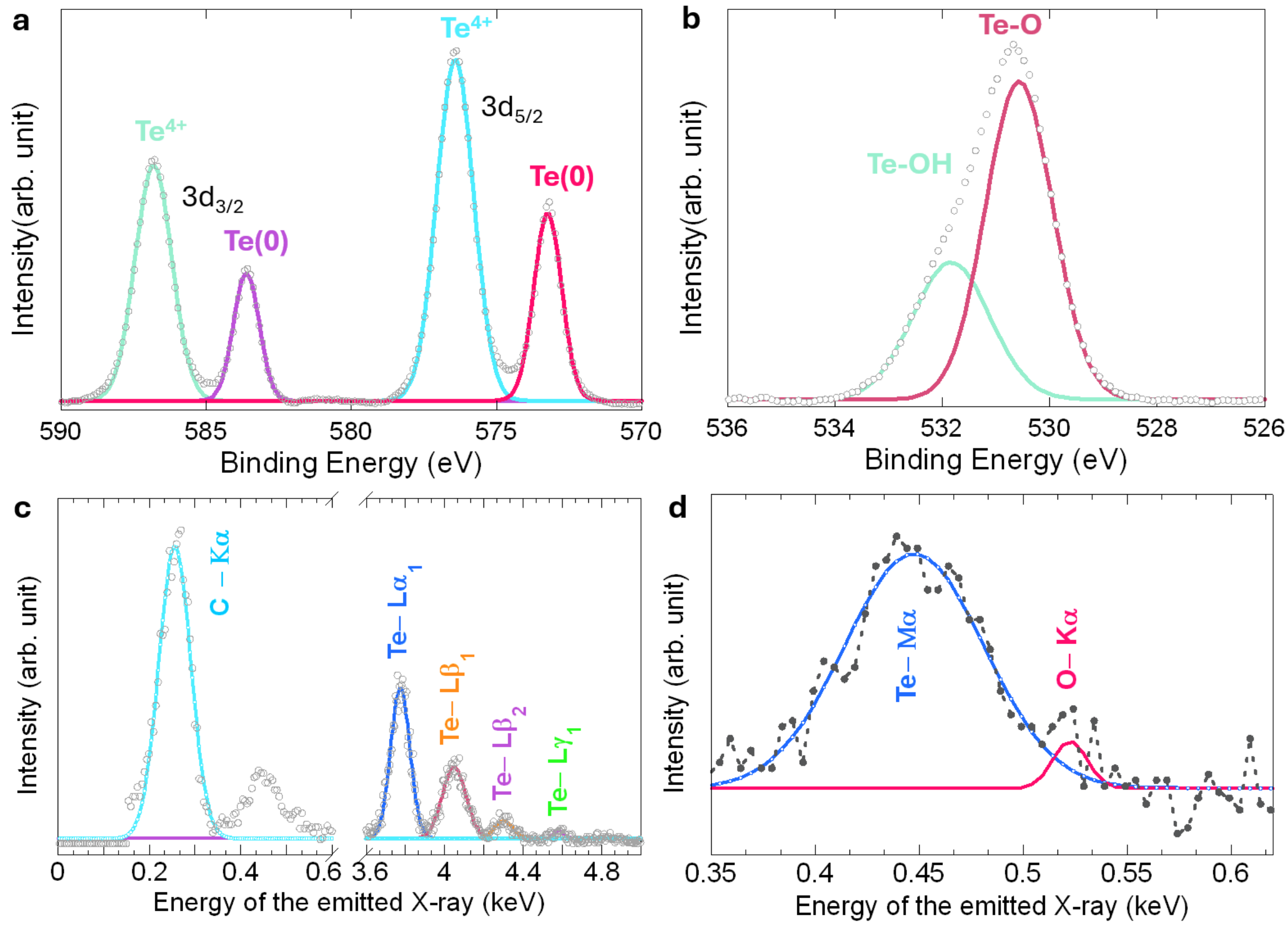

**Figure 2. (a)** High-resolution XPS spectrum of Te 3d and **(b)** O 1s. **(c)** EDS spectrum of the Te nanoparticles (TNPs), with the characteristic emission features fitted using Gaussian components. **(d)** Gaussian-fitted characteristic emission lines of Te Mα and O Kα.

Figure 2 presents complementary spectroscopic techniques for characterizing the chemical composition and oxidation states of the tellurium nanoparticles (TNPs). X-ray photoelectron spectroscopy (XPS) uses incident X-rays to eject photoelectrons, whose kinetic energies are measured to determine the elemental composition and chemical states of the near-surface region. In contrast, energy-dispersive X-ray spectroscopy (EDS) detects characteristic X-rays generated by electron-beam excitation and provides information on the elemental composition of the analysed region.

The XPS survey spectrum (Figure S4a) shows signals attributable to Te, O, and C, with no additional impurity elements detected within the measurement sensitivity. The carbon signal is consistent with the presence of carbon-containing surface species, including the organic ligands and polyvinylpyrrolidone (PVP) used during sample preparation. The high-resolution C 1s spectrum (Figure S4b) was deconvoluted into components assigned to C–C/C–H and carbonyl-containing species associated with the organic surface components.

The high-resolution Te 3d spectrum (Figure 2a) supports the coexistence of elemental Te and oxidized Te species consistent with $TeO_2$. The Te $3d_{5/2}$ and Te $3d_{3/2}$ regions were each fitted with two components. The lower-binding-energy components at 573.2 and 583.6 eV are assigned to Te(0) $3d_{5/2}$ and Te(0) $3d_{3/2}$, respectively. The higher-binding-energy components at 576.4 and 586.8 eV are assigned to Te(IV) in $TeO_2$, corresponding to the Te $3d_{5/2}$ and Te $3d_{3/2}$ levels, respectively. These assignments are consistent with reported XPS binding energies for elemental Te and $TeO_2$.[22–25] The bonding of highly electronegative O (Pauling electronegativity of 3.44) to Te (Pauling electronegativity of 2.1) at the surface pulls away electron density from the Te atoms,[26] hence, the corresponding oxidized feature appears at higher binding energy.

The O 1s spectrum (Figure 2b) was fitted with two components at 530.5 and 531.8 eV. The lower-binding-energy component is assigned to lattice oxygen in Te–O bonds within the oxide phase. [23,28] The higher-binding-energy component is attributed to surface oxygen species with hydroxyl termination at the surface (Te–OH). [27,29]

From the integrated peak areas of the Te(0) and Te(IV) components, the Te(0): Te(IV) ratio obtained from the Te $3d_{5/2}$ spectrum is 1:2.42. Similarly, the corresponding ratio obtained from the Te $3d_{3/2}$ spectrum is 1:2.65, giving an average ratio of approximately 1:2.5 (or 6:15). This indicates that, for every six unoxidized Te atoms, approximately fifteen Te atoms are present in the oxidized state on the surface. Analysis of the O 1s spectral components further reveals a Te–O: Te–O–H ratio of 1:2. Thus, of the fifteen oxidized Te atoms, approximately five are associated with Te–O bonds, while the remaining ten are associated with Te–O–H bonds.

It is important to note that the tellurium nanostructures investigated in this study exhibit a lower degree of surface oxidation than our previously reported tellurium nanostructure ensembles [8], for which the Te(0): Te(IV) ratio was 1:16, indicating a substantially higher proportion of oxidized Te atoms at the nanoparticle surfaces. The comparatively lower degree of oxidation observed here is expected to influence interparticle connectivity within the TNP ensemble. This difference is also reflected in the device dark current. Specifically, the dark current reported in our previous study was in the microampere range, whereas the devices investigated in the present study exhibit a higher dark current despite having the same device geometry and comparable thickness.

The elemental composition was further examined by EDS (Figure 2c). The measured spectrum was fitted with Gaussian components assigned to characteristic Te L-series and other elemental X-ray signals. The broad low-energy feature between approximately 0.35 and 0.55 keV was fitted using contributions assigned to Te M-series and O Kα signals (Figure 2d). The fitted peak positions and assignments are summarized in Table S1. The integrated peak-area ratio of Te $L\alpha_1$ to O Kα was approximately 81:1. After applying the Cliff–Lorimer factors[30,31], the EDS analysis yielded estimated atomic percentages of 97% Te and 3% O. These values represent the composition estimated from the analysed EDS volume and should not be interpreted as a direct measurement of the surface oxidation fraction. But this is consistent with the significantly lower oxidation of TNP. The C-Kα signal is attributed to carbon-containing surface species and/or the sample-support environment, whereas the Cu Kα signal (Figure S5) originates from the copper -carbon mesh used for electron microscopy.[32]

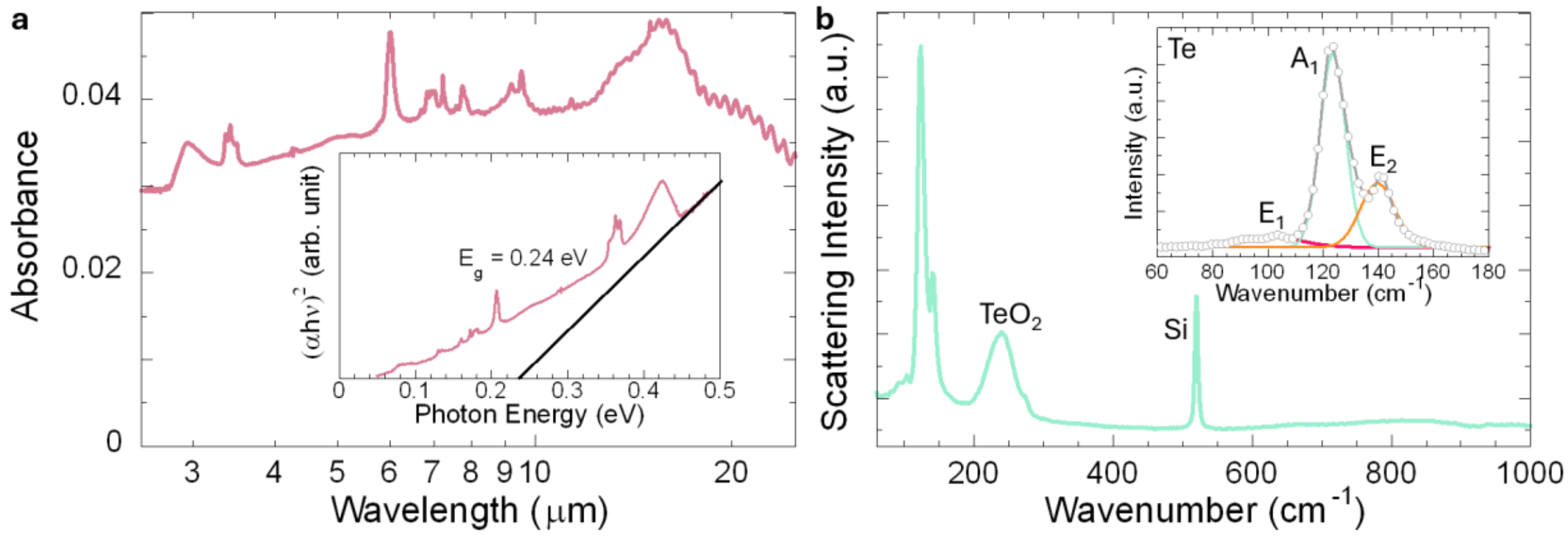

**Figure 3.** (**a**) Fourier-transform infrared (FTIR) absorption spectrum of a representative Te NP film. The inset shows the corresponding Tauc plot, yielding an estimated optical band gap of ~0.24 eV. (**b**) Raman spectrum of the Te NP film. The inset shows the Gaussian deconvolution of the Raman spectrum, highlighting the fundamental vibrational modes of tellurium.

The infrared absorption characteristics of the tellurium nanoparticle (Te NP) ensemble were investigated using Fourier-transform infrared (FTIR) spectroscopy, as shown in Figure 3a. The spectrum reveals characteristic vibrational modes associated with the molecular ligands and surface functional groups present in the Te NP ensemble. The broad absorption feature centred around 3 $\mu$m is attributed to the overlapping N–H stretching vibrations of α-methylbenzylamine and O–H stretching vibrations of surface Te–O–H groups[33], the latter being consistent with the XPS analysis. Furthermore, the absorption band at 1656 $cm^{-1}$ is assigned to the C=O stretching vibration of the carbonyl groups in the PVP backbone. A summary of the FT-IR bands and their corresponding vibrational assignments is provided in Table S2. The optical band gap, estimated from the corresponding Tauc plot (Figure 3a inset), is approximately 0.24 eV, indicating that the Te NP ensemble possesses an optical absorption threshold compatible with detection in the infrared telecommunications wavelength range.

The Raman spectrum of the Te NP ensemble (Figure 3b) exhibits characteristic vibrational features of elemental tellurium. Deconvolution of the spectrum into individual Gaussian components resolves the $E_1$, $A_1$, and $E_2$ vibrational modes at approximately 100, 123, and 140 $cm^{-1}$, respectively, as shown in the inset.[34–36] In addition, the broad feature centred around 240 $cm^{-1}$ is attributed to $TeO_2$, [37] while the sharp peak at 520 $cm^{-1}$ originates from the crystalline silicon substrate used for film deposition.[38]

To further assess the extent of surface oxidation, the integrated area of the $TeO_2$-related Raman feature at 240 $cm^{-1}$ was compared with that of the elemental Te $A_1$ mode at 123 $cm^{-1}$. The resulting area ratio ($TeO_2$:Te $A_1$) is 0.87 (Figure S6), substantially lower than the value of 2.18 obtained for the highly oxidised TNP ensemble reported in our previous study.[8] This comparison provides additional spectroscopic evidence that the Te NP ensemble investigated in the present study exhibits a lower degree of oxidation than the previously reported system.

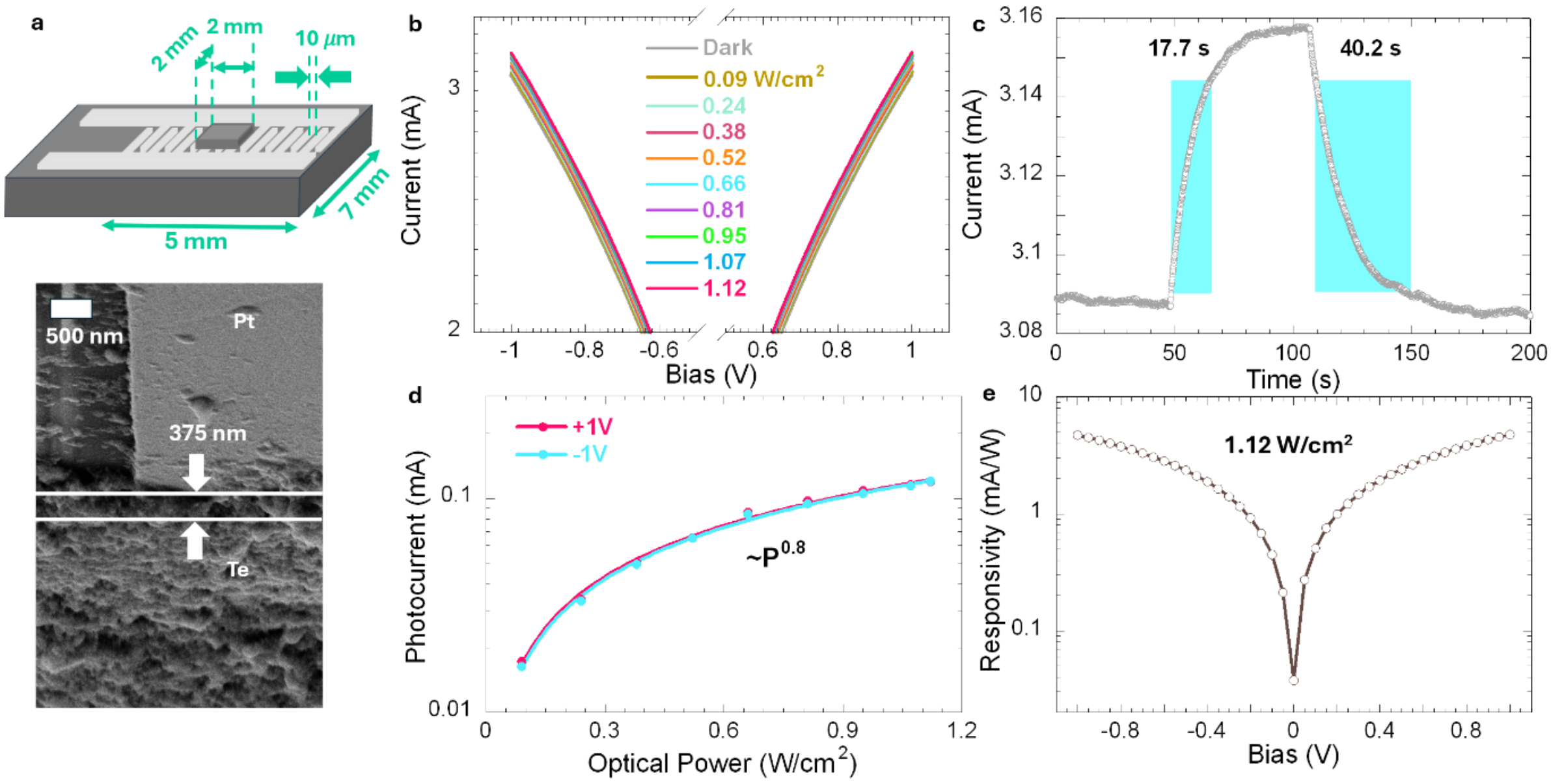

**Figure 4.** (**a**) Current–voltage (I–V) characteristics of the device under 1550 nm illumination at different incident optical power densities, as indicated in the legend. (**b**) Temporal photoresponse of the device under 1550 nm illumination with 1V bias applied. (**c**) Photocurrent as a function of incident optical power density at 1550 nm. (**d**) Bias-dependent responsivity of the photodetector under 1550 nm illumination.

The I–V characteristics of the device as shown in Figure 4a, fabricated by drop-casting the TNP ensemble onto platinum interdigitated electrodes with an electrode width and interelectrode spacing of 10 $\mu$m each, are presented in Figure 4b. The film morphology was further examined by SEM, as shown in Figure S7, which reveals interconnected TNPs and voids within the film. The I–V curves under different illumination conditions exhibit considerable overlap owing to the relatively high dark current compared with the photocurrent generated under illumination. The temporal photoresponse of the device is shown in Figure 4c. The rise and fall times were determined from the time required for the current signal to change between 10% and 90% of its total variation. The corresponding rise and fall times were estimated to be 17.7 and 40.2 s, respectively.

The photocurrent ($I_p$), defined as the difference between the current under illumination ($I_{illuminated}$) and the dark current ($I_d$), i.e., $I_p = I_{illuminated} - I_d$, exhibits a power-law dependence on the incident optical power density at 1550 nm, as shown in Figure 4d, following the relationship $I_p \propto P^{0.8}$ where $I_p$ is the photocurrent, P is the incident optical power density, and α is the power-law exponent. Similarly, the extracted exponents for illumination at 405 and 532 nm are 1.08 and 0.90, respectively (Figure S8). The nearly linear dependence of photocurrent on incident optical power across these wavelengths indicates a predominantly linear photoconductive response over the investigated power range. Such behaviour suggests that the photocurrent generation and transport processes are not strongly limited by intensity-dependent carrier recombination or trapping within the investigated optical power range. Furthermore, the bias-dependent responsivity (Figure 4e, Figure S9) exhibits approximately symmetric behaviour under positive and negative applied biases, indicating a similar photoresponse polarity and magnitude for both bias polarities. Here, the responsivity of the device is defined by the following expression:

$$R = \frac{I_p}{P_{opt}} \quad (1)$$

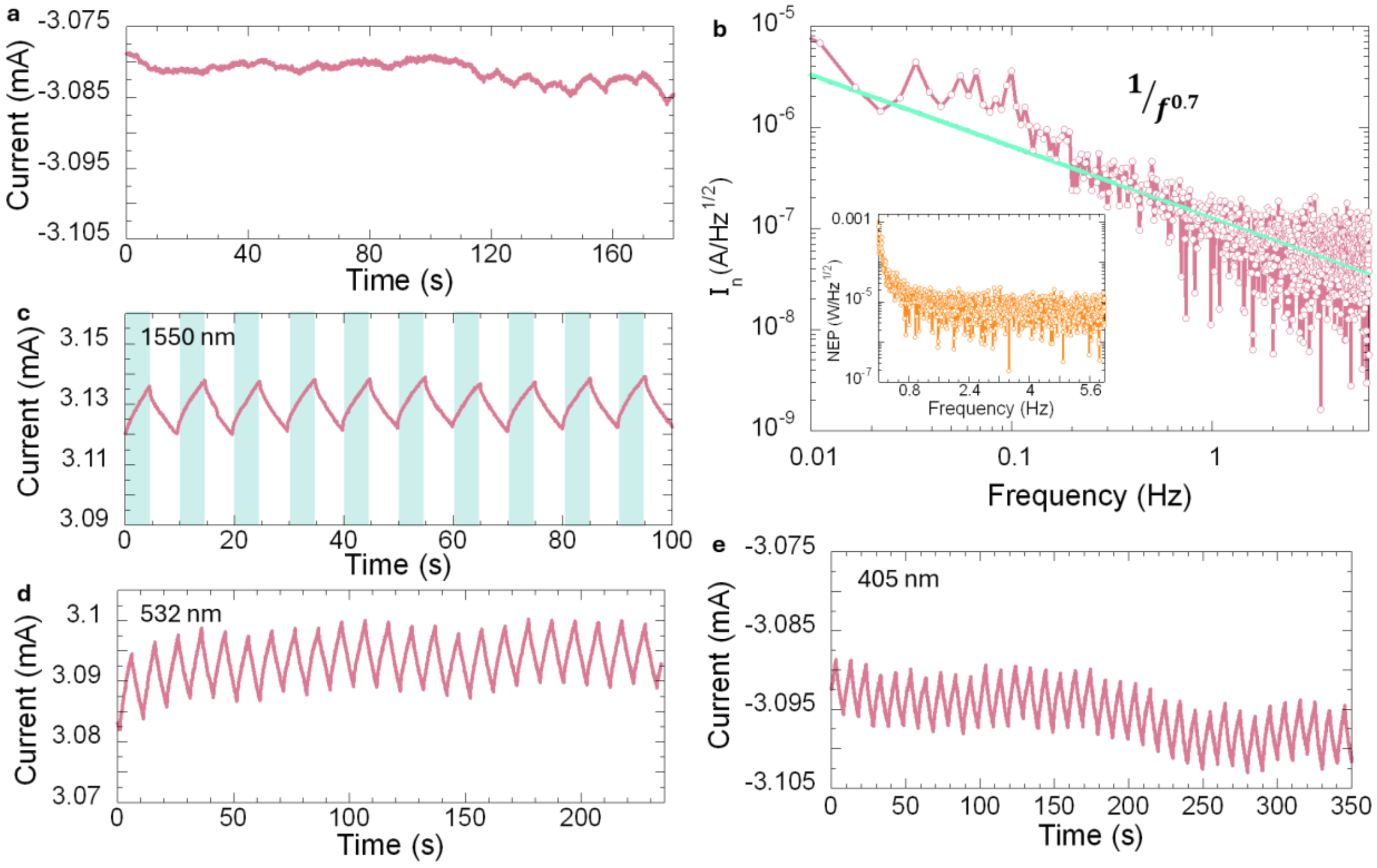

**Figure 5. (a)** Dark-current variation recorded over an extended period at an applied bias of −1 V, showing the temporal fluctuations in the dark current. **(b)** Frequency-dependent noise current spectral density and noise-equivalent power (NEP). Modulated photoresponse recorded at a chopping frequency of 0.1 Hz under illumination with **(c)** 1550 nm at an applied bias of +1 V, **(d)** 532 nm at +1 V, and **(e)** 405 nm at −1 V. The illumination periods are shaded in green in panel (c) for clarity.

The temporal fluctuations in the dark current were monitored over a period of 180 s at an applied bias of −1 V to evaluate the low-frequency noise characteristics of the device, as shown in Figure 5a. The corresponding noise-current spectral density, derived from the power spectral density (Figure S10), is presented in Figure 5b and reveals a pronounced contribution from flicker noise. The frequency dependence of the noise current follows a power-law relationship, $I_n \alpha\, 1/f^{0.7}$ where (f) is the noise frequency. Owing to the acquisition time and measurement bandwidth of the experimental setup, the noise spectrum could only be reliably evaluated up to approximately 6 Hz. Consequently, the measured spectrum primarily reflects the low-frequency noise characteristics of the device, where flicker noise is expected to make a substantial contribution. The noise-equivalent power (NEP) was estimated from the noise-current spectral density and the device responsivity under 1550 nm illumination, according to

$$NEP = \frac{I_n}{R} \tag{2}$$

where R is the responsivity. The corresponding NEP spectra for the two measurement conditions are presented in Figure S11a-b. The NEP approaches a plateau of $10^5$ W/Hz$^{1/2}$ towards the higher-frequency end of the measured range.

For comparison, the theoretical shot-noise and thermal-noise current spectral densities were estimated using the following expressions, respectively:

$$I_{n,shot} = \sqrt{2eI_d} \tag{3}$$

$$I_{n,thermal} = \sqrt{4k_BT/R_{device}} \tag{4}$$

respectively, where e is the elementary charge, $I_d$ is the dark current, $k_B$ is the Boltzmann constant, T is the absolute temperature, and $R_{device}$ is the electrical resistance of the device. The estimated shot-noise and thermal-noise current spectral densities are 31.47 and 7 pA/Hz$^{1/2}$, respectively. However, the limited frequency range accessible with the present measurement system prevented the experimental evaluation of the noise spectrum in the higher-frequency regime, where the contribution from flicker noise may become less significant. Therefore, the specific detectivity was estimated using the conventional shot-noise-limited expression, assuming that shot noise represents the dominant noise contribution under the relevant operating conditions.[39] This value should consequently be regarded as a theoretical shot-noise-limited estimate rather than a detectivity determined directly from the experimentally measured total noise.

The observed low-frequency flicker noise may originate from fluctuations in carrier number and mobility within the heterogeneous Te nanostructure network. In particular, structural distortions and lattice mismatch at the Te/$TeO_2$ interfaces may introduce localized defect states that facilitate carrier trapping and detrapping. Fluctuations in the occupation of these states, together with the associated variations in carrier mobility and interparticle charge-transfer pathways, may contribute to the observed low-frequency noise.[40–42]

The stability and reproducibility of the photoresponse were investigated over multiple illumination cycles using light sources of different wavelengths at a chopping frequency of 0.1 Hz, as shown in Figure 5c-e. The device exhibits repeatable photoresponse over successive illumination cycles, demonstrating its ability to operate under periodically modulated optical excitation.

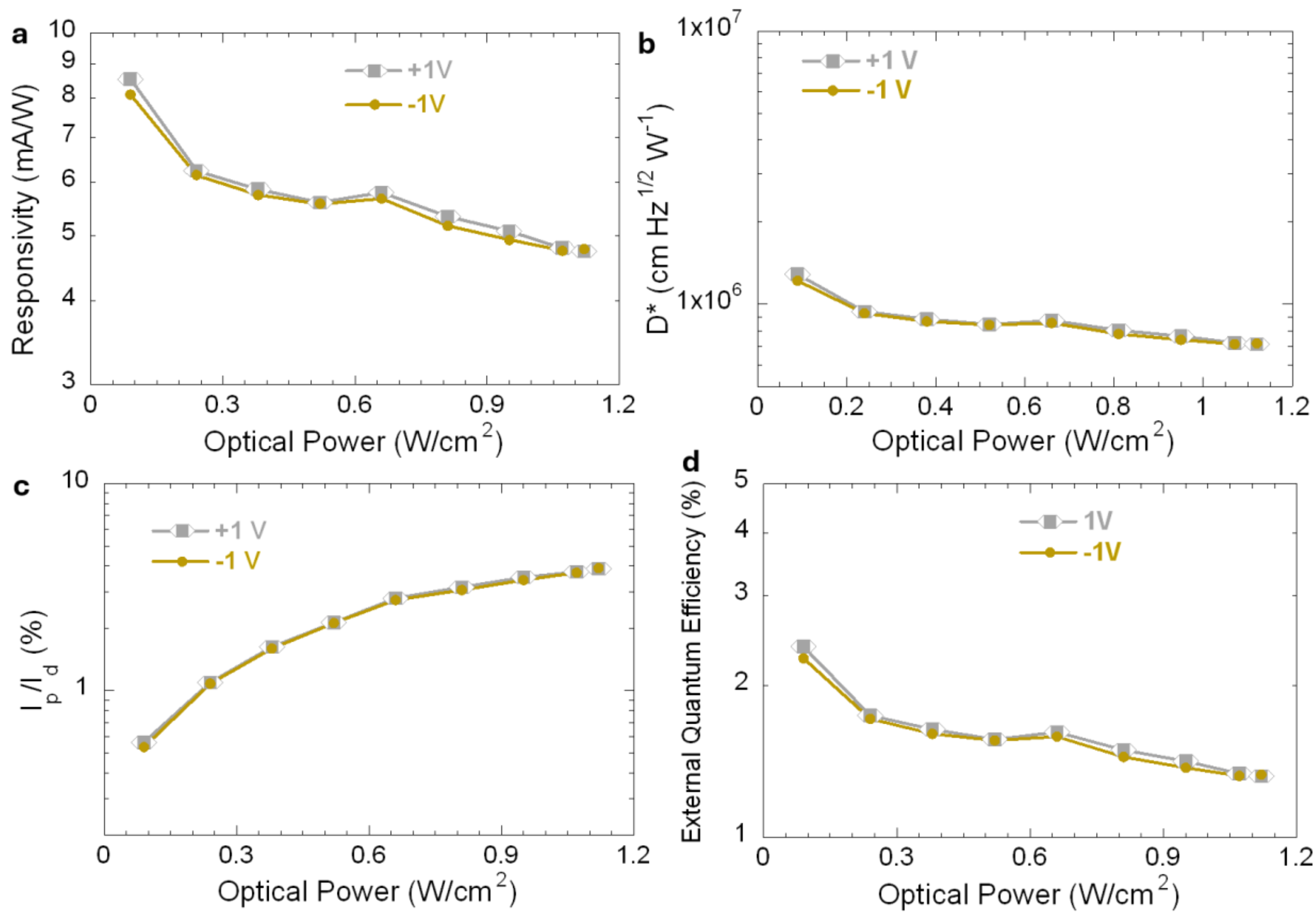

**Figure 6.** Device performance metrics under 1550 nm illumination as a function of incident optical power density: (**a**) Responsivity, (**b**) specific detectivity, (**c**) photocurrent-to-dark-current ratio, and (**d**) external quantum efficiency (EQE).

The responsivity of the device, calculated at applied biases of +1 and −1 V, is presented in Figure 6a. The responsivity decreases from approximately 9 to 5 mA/W with increasing incident optical power density. The corresponding shot-noise-limited specific detectivity[39] was estimated to be approximately $10^6$ cm $Hz^{1/2}$ $W^{-1}$ (Figure 6b). The photocurrent-to-dark-current ratio, $I_p/I_d$, is another important figure of merit, as it represents the photoinduced current relative to the background dark current. This ratio increases with incident 1550 nm optical power density, reaching approximately 4% at 1.12 W/cm$^2$, as shown in Figure 6c. The maximum external quantum efficiencies (EQEs), calculated using the following expression, were found to be 2.4%, 4.03%, and 2.14% under illumination at 1550, 532, and 405 nm, respectively (Figure 6d, S13b and S14b). The EQE decreases with increasing optical power density, consistent with the observed reduction in responsivity.

$$EQE = \frac{Rch}{e\lambda} \quad (5)$$

where R is the responsivity, h is Planck's constant, c is the speed of light, e is the elementary charge, and λ is the incident wavelength.

The modest detectivity value compared to earlier studies, as shown in Table S3, may be associated with the relatively high dark current, in the milliampere range, arising from the interconnected Te nanoparticle (TNP) network. Consequently, the photoinduced current represents a relatively small contribution to the total device current. Although controlled oxidation of the TNP film may improve interparticle charge transport and enhance the photocurrent generated at a given optical power, the high background dark current limits the photocurrent-to-dark-current ratio and, consequently, the detectivity.[43,44] Thus, improvements in charge transport and responsivity do not necessarily translate into a corresponding enhancement in detectivity. The corresponding device figures of merit under 532 and 405 nm illumination are presented in Figures S12 and S13, respectively. The maximum responsivities achieved under illumination at 1550, 532, and 405 nm were 8.5, 14.4, and 10.1 mA/W, respectively. Correspondingly, the maximum specific detectivities were estimated to be 1.3 × $10^6$ , 6.87 × $10^7$, and 1.3 × $10^7$ Jones, respectively. Although the responsivity, specific detectivity, and EQE values are modest compared with those of state-of-the-art tellurium-based photodetectors (Table S1), this work highlights the potential of surface oxide passivation as a means of engineering interparticle charge transport and improving the performance of solution-processed tellurium-based photodetectors for short-wave infrared (SWIR) detection.

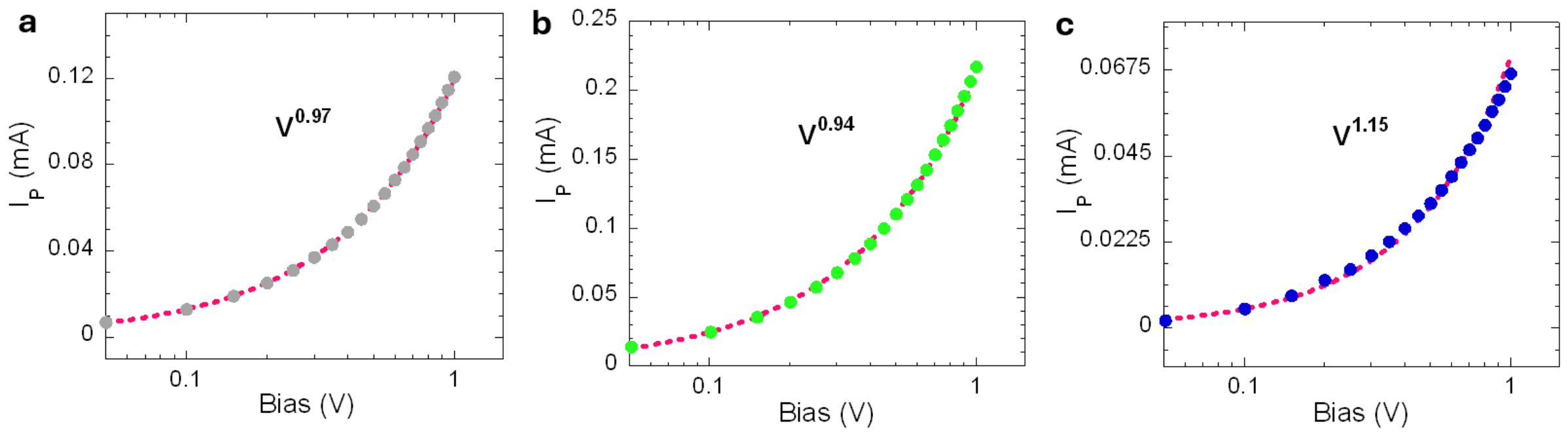

**Figure 7.** Power-law dependence of photocurrent on applied bias at different illumination conditions: **(a)** 1550 nm (1.12 W/cm$^2$), with an exponent of 0.97; **(b)** 532 nm (1.42 W/cm$^2$), with an exponent of 0.94; and **(c)** 405 nm (0.2 W/cm$^2$), with an exponent of 1.15, respectively, as indicated in the plots. Pink dashed lines represent the fitted curves.

The bias dependence of the photocurrent was further examined using a power-law relationship, $I_p = V^n$, where n is the exponent. The extracted exponents were 0.97, 0.94, and 01.15 under illumination at 1550, 532, and 405 nm, respectively. The values close to unity indicate an approximately linear increase in photocurrent with applied bias across the investigated spectral range. The bias-dependent photocurrent follows an approximately linear power-law relationship, with exponents of 0.97, 0.94, and 1.15 under 1550, 532 and 405 nm illumination, respectively (Figure 7). These values indicate a broadly near-linear dependence of photocurrent on applied bias, consistent with field-assisted facile carrier transport and collection through the lateral interdigitated-electrode geometry.[45] Furthermore, the photocurrent exhibits an approximately linear dependence on optical power density over the investigated range (Figures 4d and S9), indicating that the photoresponse scales nearly proportionally with the incident photon flux. This behaviour suggests that photocarrier generation and subsequent collection occur without pronounced saturation over the measured illumination range. The linear current–voltage characteristics observed under both dark and illuminated conditions (Figure S14) further suggest near-ohmic electrical behaviour at the device level, with no pronounced rectification over the investigated bias range. Hence, these observations are consistent with effective photocarrier generation and extraction through the Te nanoparticle network, facilitated by favourable electrical contact with the Pt electrodes. Although these characteristics support efficient charge collection, they do not independently establish negligible carrier recombination or completely barrier-free transport at the Te–Pt interface.

CONCLUSION:

In conclusion, we demonstrate that surface oxidation plays an important role in governing the optoelectronic properties of solution-processed tellurium nanoparticle (TNP) ensembles. Gradual ambient-air exposure produces a comparatively lower relative contribution of oxidized Te species than the previously investigated forced-air-injection approach, while retaining coexisting elemental Te and α-$TeO_2$. The resulting nanoparticle ensemble exhibits photodetection across 405, 532, and 1550 nm, with responsivities of 10.1, 14.4, and 8.5 mA/W, respectively. However, the relatively high dark current and pronounced low-frequency flicker noise limit the specific detectivity. These findings highlight the importance of balancing surface oxidation and interparticle electrical connectivity to regulate charge transport in nanoparticle assemblies. Further control over oxidation and interparticle interfaces could provide a route to improving the performance of solution-processed Te-based photodetectors for SWIR applications.

METHOD SECTION:

**Synthesis of Tellurium Nanoparticles:**

A thermodynamically controlled gradual oxidation of telluride precursor was considered for synthesizing the tellurium nanoparticles with optimum oxide layer thickness. In conventional colloidal semiconductor quantum dots, such as CdSe/ZnS, CdSe/CdS, and $CuInS_2$/CdS, the structural integrity, shell thickness, and electronic band alignment of the shell play important roles in determining surface-defect passivation and charge transfer from the quantum-dot core to external acceptors or electrodes. Depending on the band alignment and shell thickness, the shell can either facilitate charge separation through carrier delocalization and defect passivation or impede carrier extraction by introducing an additional tunnelling barrier.[46–49] In assembled quantum-dot films, efficient carrier collection further requires adequate electronic coupling between neighbouring nanoparticles to establish continuous charge-transport pathways to the electrodes.[50] Therefore, the method followed here uses slow oxidation with probable thermodynamically controlled reduction of telluride in the presence of amine and PVP polymer. It was hypothesized that the amine is used to coordinate with tellurium, and the PVP polymer helps the tellurium chain to form a long helical chain. [51–54]

In an $N_2$-filled glovebox, α-methylbenzylamine (1.5 mL, 13 mmol) and polyvinylpyrrolidone (PVP, 200 mg; weight-average molecular weight, ($M_w$ = 40,000 g $mol^{-1}$) were added to a 5 mL glass vial equipped with a magnetic stir bar. The mixture was stirred at 1000 rpm until PVP was completely dissolved in the amine. Subsequently, bis(trimethylsilyl)telluride ($(TMS)_2Te$) 100 $\mu$L, 0.354 mmol) was added to the solution, resulting in the development of a pink coloration that gradually intensified with time. The reaction mixture was maintained inside the glovebox for 24 h.

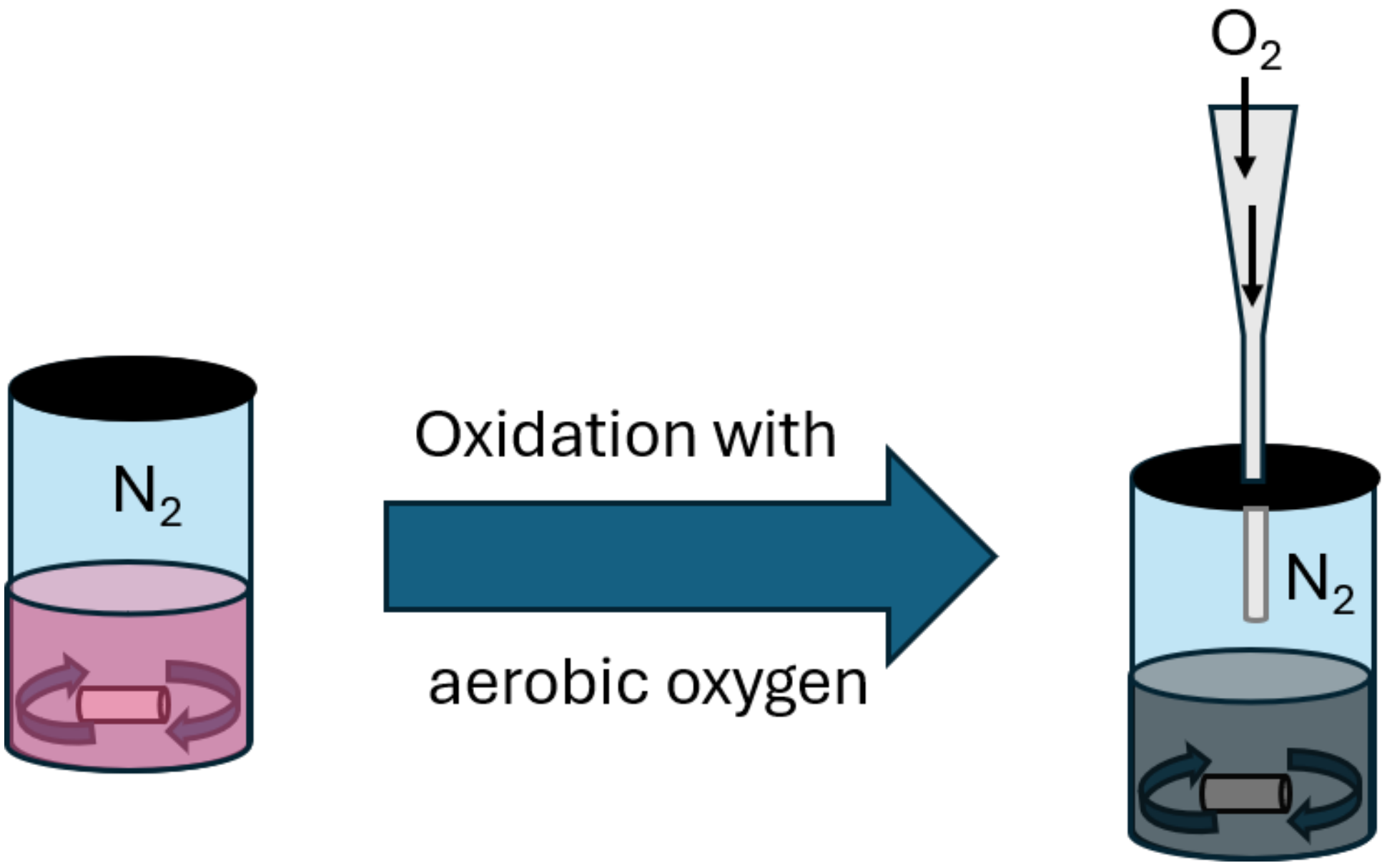

**Figure 8**. Schematic of the oxidative pathway of tellurium nanoparticle synthesis.

After 24 h, the vial was removed from the glovebox and immediately sealed with an air-tight rubber septum. A 21-gauge needle was inserted through the septum to allow controlled exposure of the reaction mixture to atmospheric oxygen as shown in the schematic of Figure 8. This exposure initiated the oxidation of telluride species, facilitating the formation of elemental Te through a two-electron oxidation process

$$2Te^{2-} + O_2 + 2H_2O = 2Te(0) + \ 4OH^-$$

In the presence of α-methylbenzylamine and PVP, the resulting Te species underwent nucleation and subsequent growth to form Te nanocrystals. The oxidation process was allowed to proceed under ambient conditions for seven days, during which the pink coloration gradually disappeared (Figure S15), indicating the conversion of the telluride precursor into Te nanocrystals. Similarly, $TeO_2$ formation happens after the surface tellurium is exposed to oxygen.

$$Te \ + O_2 = \ TeO_2$$

The gradual exposure of the reaction mixture to atmospheric oxygen was intended to moderate the oxidation of the telluride precursor and the subsequent supply of Te-containing growth species. Following the initial nucleation event, the relatively slow generation of these species may favour the continued growth of existing Te nanocrystals over rapid, uncontrolled particle formation after the burst nucleation process when the monomer(just produced tellurium atoms from oxidation of $Te^{2-}$) is higher than the critical concentration according to LaMer's model of nucleation.[55–57] This gradual oxidation may also allow the surface oxide to develop progressively around the Te-rich cores. We propose that structural disorder and defect-associated electronic states within the resulting $TeO_2$-rich interfacial layer could influence carrier transport across the core–shell interface, potentially providing intermediate states for thermally assisted hopping or trap-assisted tunnelling. In this picture, photogenerated carriers are first produced in the semiconducting Te cores and may subsequently traverse the thin oxide-rich interface. However, the extent to which such states facilitate transport, rather than act as carrier traps or barriers, requires further experimental verification.

Following the oxidation process, the crude reaction mixture was washed three times with acetone to remove residual organic ligands and other soluble reaction by-products. For purification, 1 mL aliquots of the dispersion were transferred into multiple 2 mL Eppendorf tubes. Each aliquot was briefly sonicated for 5 s to minimize potential degradation of the Te nanocrystals, followed by centrifugation at 4000 rpm for 6 min. The supernatant was carefully decanted, and the resulting precipitate was redispersed in fresh acetone using brief sonication. This washing and centrifugation procedure was repeated twice more. The purified Te nanocrystals were subsequently dispersed in chloroform for further characterization and device fabrication.

**Device Fabrication and Characterisation:**

For device fabrication, 10 mg of the purified Te nanocrystals was dispersed in 1 mL of chloroform and briefly sonicated for 5 s to obtain a homogeneous dispersion. The resulting dispersion was drop-cast onto glass substrates patterned with Pt interdigitated electrodes (IDE), having an electrode width and interelectrode spacing of 10 $\mu$m each. The deposited films were allowed to dry under ambient conditions for 24 h before optoelectronic characterization.

The two terminals of the Pt interdigitated electrodes were electrically contacted using probe manipulators connected to a Keithley 2450 source measurement unit (SMU). The device was illuminated using light sources with wavelengths of 1550, 532, and 405 nm. The incident light was directed onto the device via reflection from a coated silver mirror. The optical power density at the sample position was determined by placing a power meter at the device location and measuring the illuminated spot size. The power density was calculated by dividing the measured optical power by the area of the circular illumination spot. The spot diameter was maintained larger than the lateral dimensions of the device to ensure uniform illumination of the entire active area. Accordingly, the optical power incident on the device was estimated by multiplying the measured power density by the device's active area. Since the power density was measured at the sample position, optical losses associated with reflection from the silver mirror were inherently accounted for. The incident optical power was varied using neutral-density (ND) filters.

**X-ray Diffraction (XRD)**

X-ray diffraction (XRD) measurements were carried out using a PANalytical Empyrean diffractometer equipped with a Cu Kα radiation source (λ ≈ 1.5406 Å). The synthesized tellurium nanostructures were deposited onto clean glass slides by drop-casting and subsequently dried before measurement. The X-ray source was operated at 45 kV and 30 mA. Upon interaction with the periodically arranged atomic planes of the crystalline material, the incident X-rays undergo scattering, producing constructive interference at specific diffraction angles. This condition is described by Bragg's law, ($n\lambda = 2d\ sin\theta$), where n is the diffraction order, $\lambda$ is the incident X-ray wavelength, d represents the interplanar spacing, and $\theta$ is the Bragg angle. The diffracted intensity was measured as a function of the scattering angle $2\theta$, and the resulting diffraction patterns were examined to determine the crystalline phases and structural features of the synthesized nanostructures.

**Transmission Electron Microscopy (TEM)**

Transmission electron microscopy (TEM) measurements were performed using a Titan 300 Themis microscope. Samples were prepared by depositing the nanostructure dispersion onto carbon-coated copper (Cu-C) grids, followed by drying. Elemental analysis and spatially resolved elemental mapping were conducted using a Bruker Quantax X-FLASH energy-dispersive X-ray spectroscopy (EDS) system equipped with a silicon drift detector (SDD). During EDS analysis, characteristic X-rays emitted from the sample enter the silicon detector and generate electron–hole pairs through energy deposition within the detector volume. An applied electric field drives the generated charge carriers towards the collecting anode, producing a voltage pulse. The amplitude of this pulse is proportional to the energy of the incident X-ray photon, enabling elemental identification based on characteristic X-ray energies. The detector electronics also analyse the pulse shape, including its rise time, which is influenced by the position at which the X-ray is absorbed within the detector.

**Scanning Electron Microscopy (SEM)**

Surface morphology was investigated using an Ultra 55 scanning electron microscope (Zeiss). For sample preparation, the synthesized nanostructures were deposited onto silicon wafer substrates and dried before imaging. Measurements were conducted under vacuum at a chamber pressure of $1.63 \times 10^{-5}$) mbar. Secondary-electron images were acquired using an Everhart–Thornley detector, which collects secondary electrons emitted from the sample surface under electron-beam irradiation to reveal morphological and topographical features.

**X-ray Photoelectron Spectroscopy (XPS)**

X-ray photoelectron spectroscopy (XPS) measurements were performed using an AXIS Ultra DLD spectrometer (Kratos Analytical) under a vacuum pressure of $5.2 \times 10^{9}$ Torr. Photoelectrons were generated using an Al Kα X-ray source (photon wavelength ≈ 8.34 Å). A charge-neutralization system was employed during data acquisition to compensate for surface charging effects by supplying a flux of low-energy electrons to the sample.

**Infrared Absorption Spectroscopy**

Infrared absorption measurements were conducted using a PerkinElmer Fourier-transform infrared (FTIR) spectrometer in transmission mode. The sample was prepared as a pellet by mixing the synthesized nanostructures with dry potassium bromide (KBr). Spectral measurements were performed over the wavelength range of 2.5–25 $\mu$m. The spectrometer operates on the principle of a Michelson interferometer, in which the movement of a mirror generates a varying optical path difference between two components of the infrared beam. The resulting interferogram contains contributions from the different infrared frequencies present in the incident radiation. Background and sample interferograms were recorded separately using a pure KBr pellet and a sample-containing KBr pellet, respectively. Fourier transformation of these interferograms converted the measured signals from the optical path difference domain into the wavenumber domain. The absorbance spectrum was subsequently obtained by comparing the sample spectrum with the KBr background, thereby correcting for contributions from atmospheric moisture, gases, and other background components in the optical path.

**Raman Spectroscopy**

Raman spectroscopy was performed using a Horiba Raman spectrometer with a 532 nm laser as the excitation source. The synthesized tellurium nanostructures were deposited onto a polished silicon (Si) wafer by drop-casting and dried before measurement. The focused laser beam illuminated the sample, and the resulting scattered radiation was collected through the microscope optics. Rayleigh scattering was suppressed using an optical rejection filter to isolate the inelastically scattered Raman signal. The collected light was then spectrally resolved using a diffraction grating and detected with a charge-coupled device (CCD) detector. Raman shifts were expressed in $cm^{-1}$ and determined from the difference between the wavenumbers of the incident laser and the scattered radiation.

**Acknowledgements:** The author gratefully acknowledges Anshu Pandey for helpful discussions and Ambresh M for assistance with transmission electron microscopy. Financial support from the Government of India through the Prime Minister's Research Fellowship (PMRF), Grant No. PM/MHRD-19-14871.03, is gratefully acknowledged. The author also thanks the administration and scientific community of the Indian Institute of Science for their support.

**Author Contributions:** The author independently conceived and conducted the experiments, collected and analysed the data, and prepared and revised the manuscript. The author acknowledges the use of AI solely for language editing and improving the clarity and readability of the manuscript. The final version was carefully reviewed and approved by the author, who takes full responsibility for its content.

Supporting Information

# Controlled Oxidation of Tellurium Nanoparticles for Telecom Detection

Biman Jana*

Indian Institute of Science, Bangalore – 560012, India

*Corresponding Author, Email: biman.jana1050@gmail.com

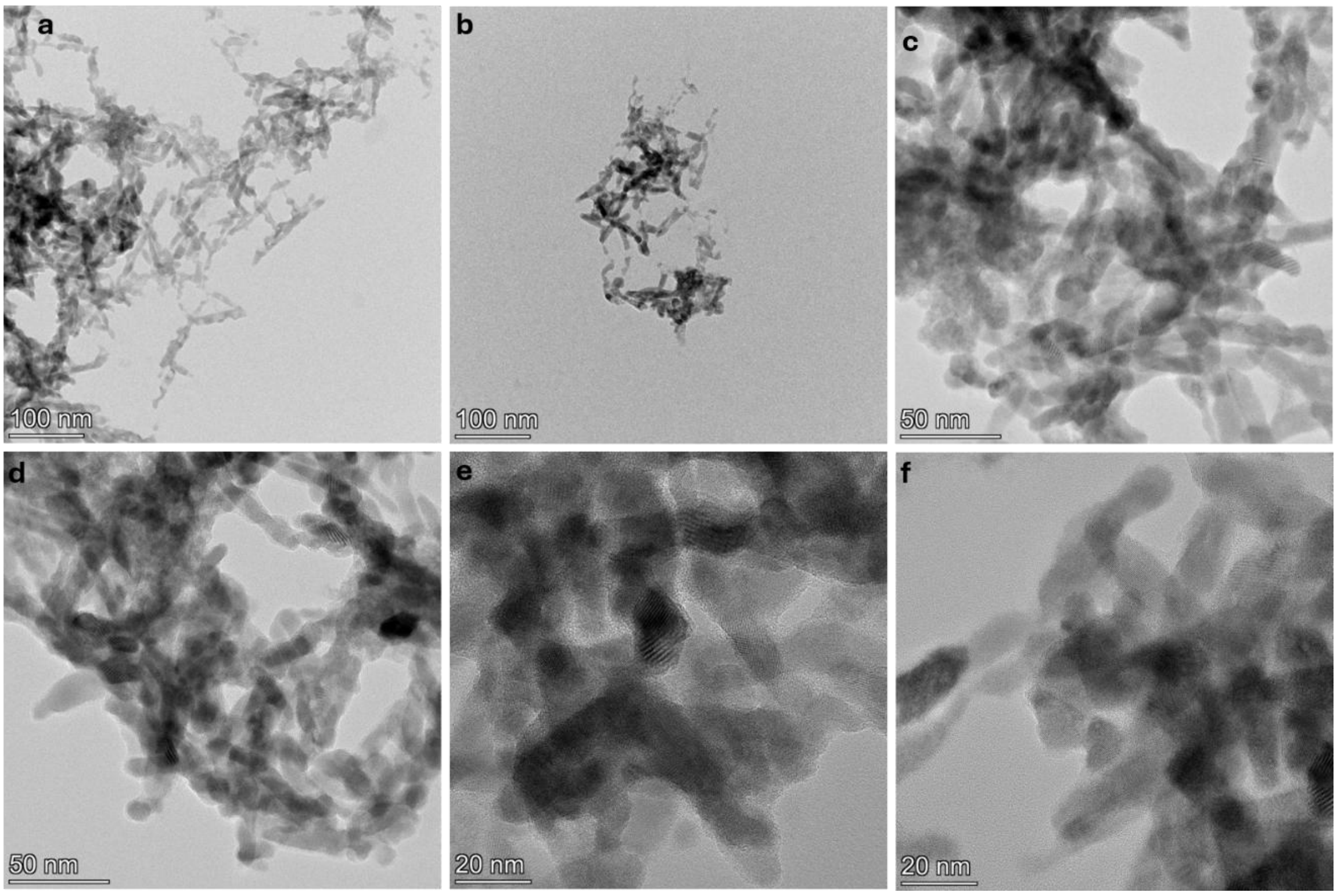

**Figure S1**. Bright-field image of Te nanoparticles.

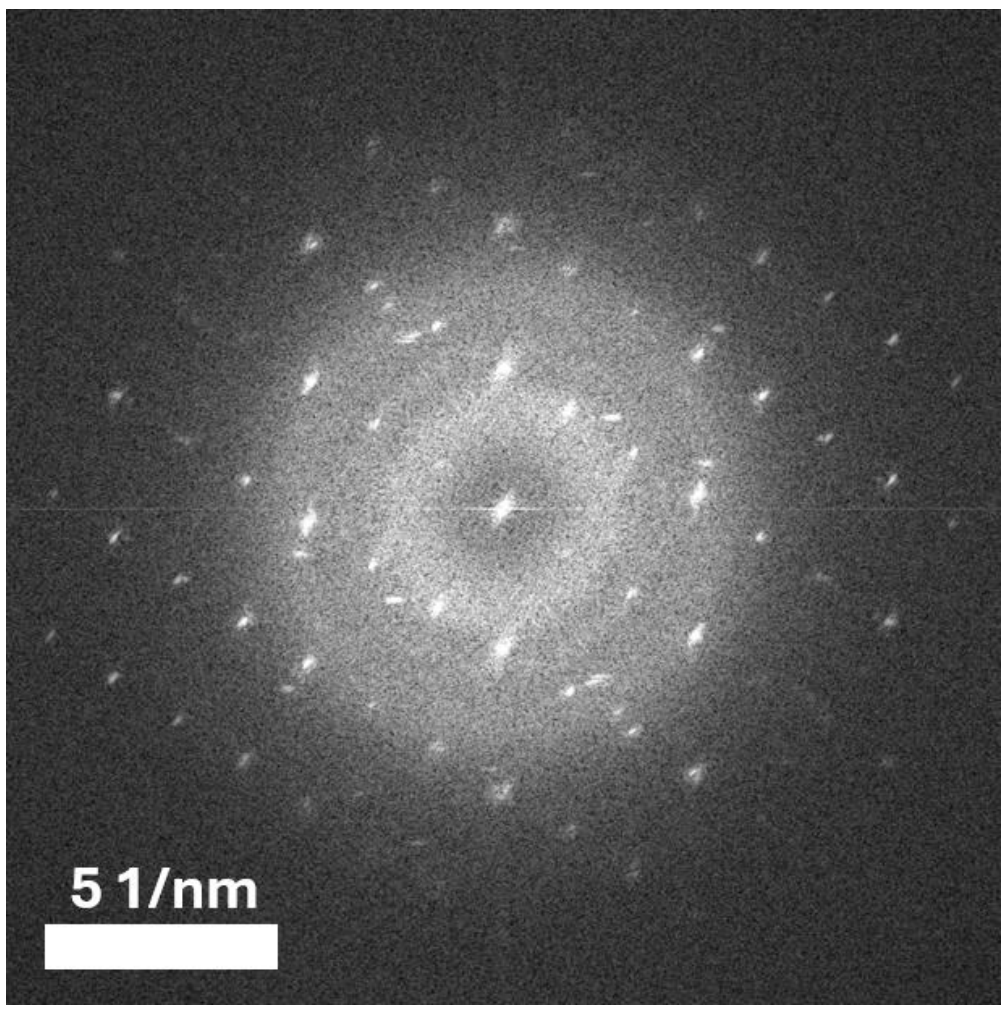

**Figure S2**. SAED pattern of the scanned region presented in Figure 1b of the main manuscript.

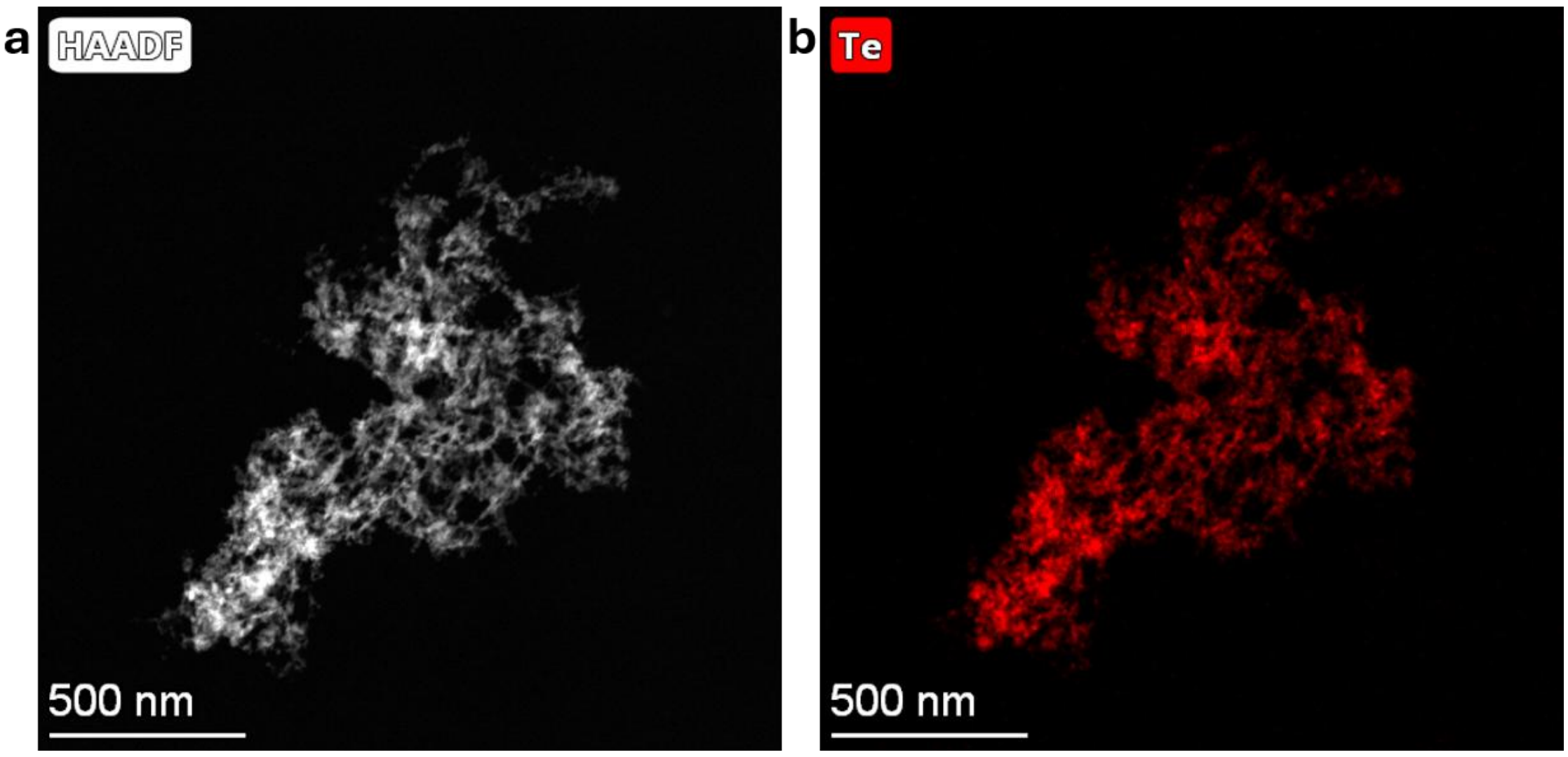

**Figure S3**. **(a-b)** HAADF image of the TNPs and corresponding elemental mapping for Te.

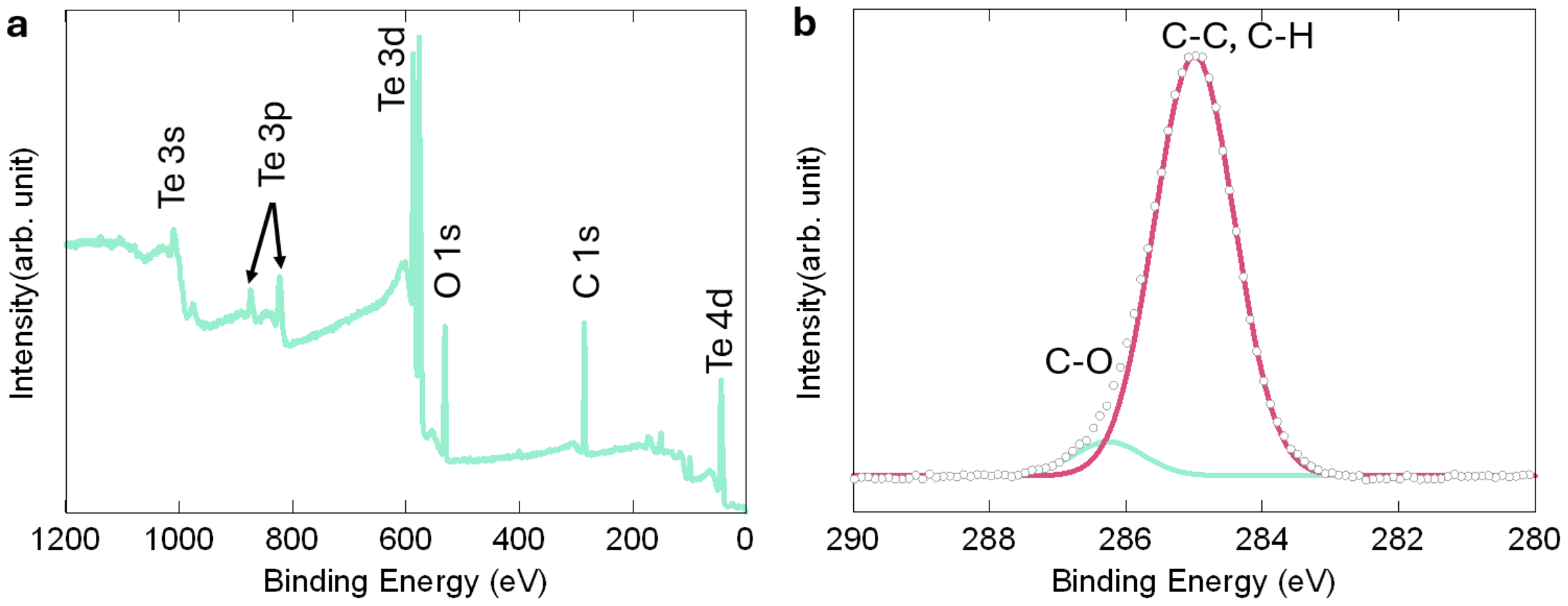

**Figure S4**. **(a)** XPS survey spectrum **(b)** for C 1s of adventitious carbon.

The XPS survey spectrum (Figure S4a) of the tellurium (Te) nanoparticle sample confirms the presence of tellurium at the sample surface. Signals corresponding to carbon and oxygen are also observed, which may originate from adventitious surface contamination, adsorbed species, surface oxidation, and organic ligand moieties. No additional elements are detected within the sensitivity of the survey scan. High-resolution Te 3d spectra are presented in Figure 3a-b. Adventitious carbon XPS spectra (Figure S4b) show the C-C $sp^3$ feature appears at 284.9 eV; hence all other XPS data have been adjusted by 0.1 eV because the standard binding energy of the photoelectron for the C-C sp3 feature is 284.8 eV

**Table S1.** EDS peaks and characteristic feature assignment:

| No. | Peak Position(keV) | Width (keV) | FWHM (keV) | Characteristic Feature |
|---|---|---|---|---|
| 1 | 0.25 | 0.07 | 0.083 | C-Ka |
| 2 | 0.44 | 0.07 | 0.078 | Te-Ma |
| 3 | 0.52 | 0.02 | 0.019 | O-Ka |
| 4 | 3.77 | 0.098 | 0.115 | Te-$La_1$ |
| 5 | 4.05 | 0.12 | 0.142 | Te-$Lb_1$ |
| 6 | 4.3 | 0.092 | 0.108 | Te-$Lb_2$ |
| 7 | 4.58 | 0.053 | 0.062 | Te-$Lg_1$ |
| 8 | 8.04 | | | Cu-Ka (From TEM grid) |

The characteristic X-ray features are assigned to Te, O, along with the C and Cu[1] . The additional C and Cu features are due to the Cu-C mesh, on which the sample was drop-cast for TEM experiments.

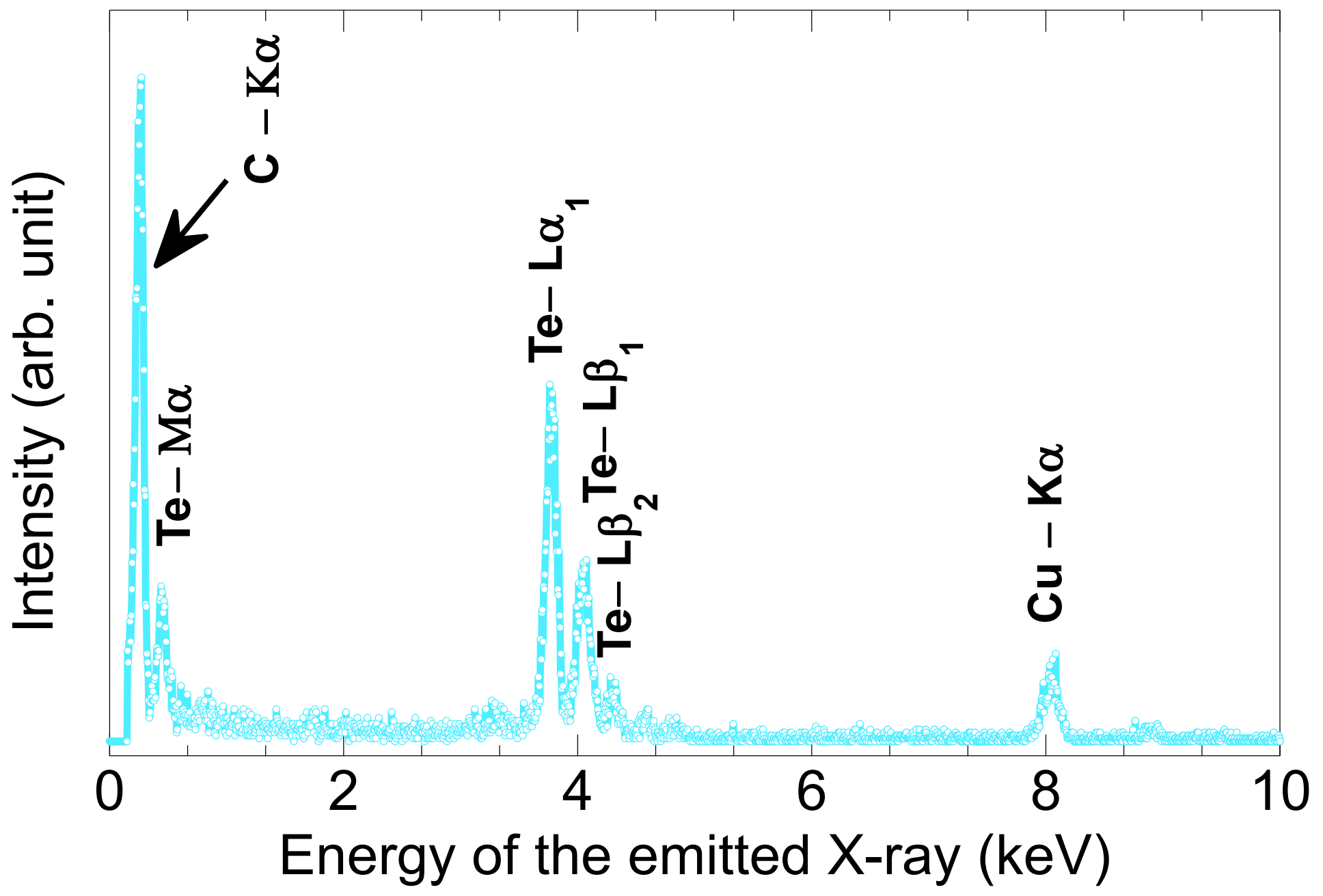

**Figure S5**. Full-scale EDS spectra of the sample with Cu-Kα feature at 8.04 keV.

**Table S2**. Assignment of FT-IR peaks to corresponding vibrational features.

| Wavenumber ($cm^{-1}$) | Corresponding Vibration |
|---|---|
| 3400 | O-H, N-H stretching |
| 2933 | C-H stretching |
| 1656 | PVP carbonyl group (C=O) stretching |
| 1457 | $CH_2$, $CH_3$ deformation |
| 1384 | C-H deformation |
| 1281 | C-N stretching of PVP and amine |
| 1091 | C-C, C-N, C-O fingerprint region |
| 1050 | C-C, C-N, C-O fingerprint region |
| 648 | Te-O |

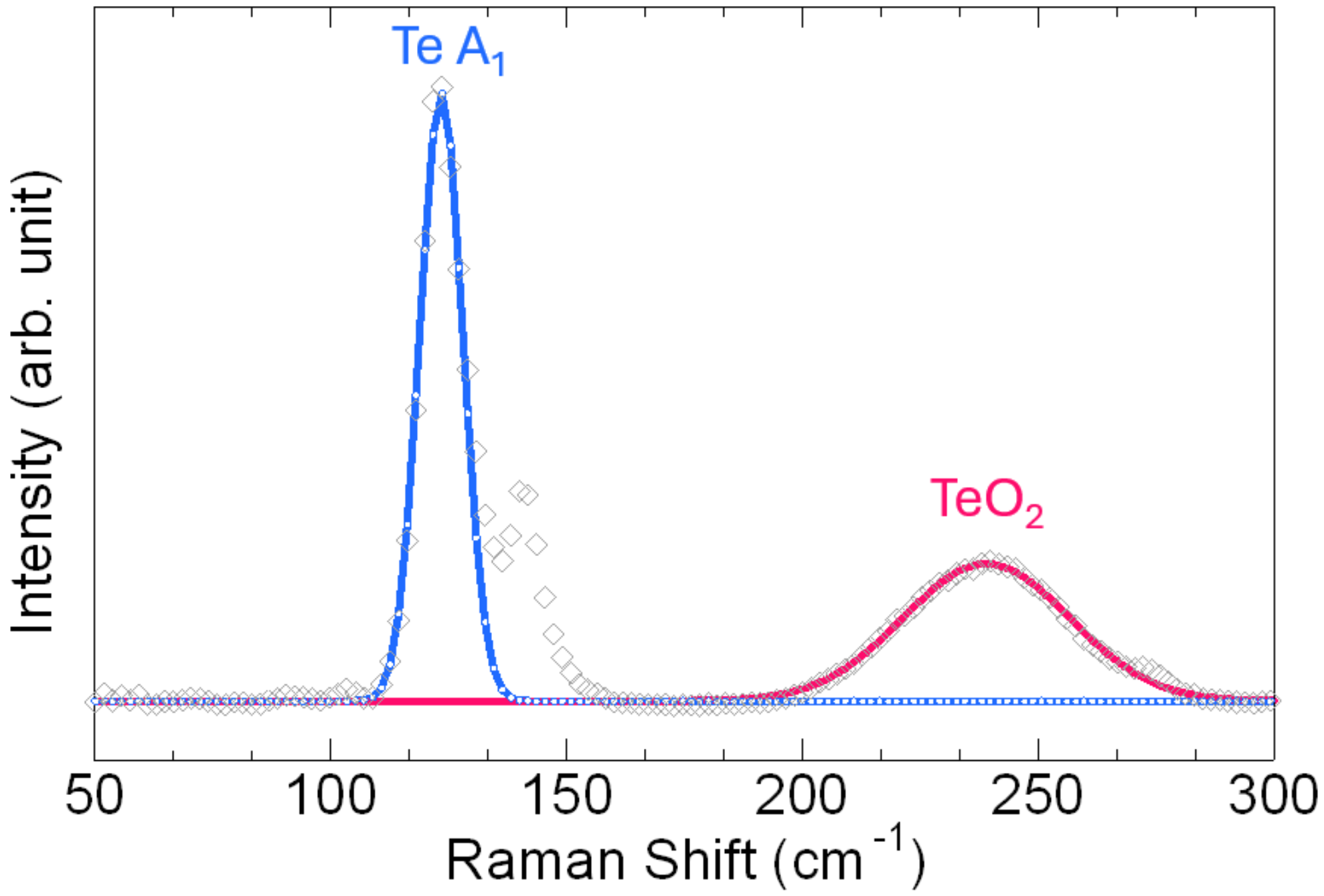

**Figure S6**. Raman spectra of the tellurium nanoparticles with Gaussian fitting of the most intense Raman feature of Te (i.e. $A_1$ mode) and the feature for $TeO_2$.[2]

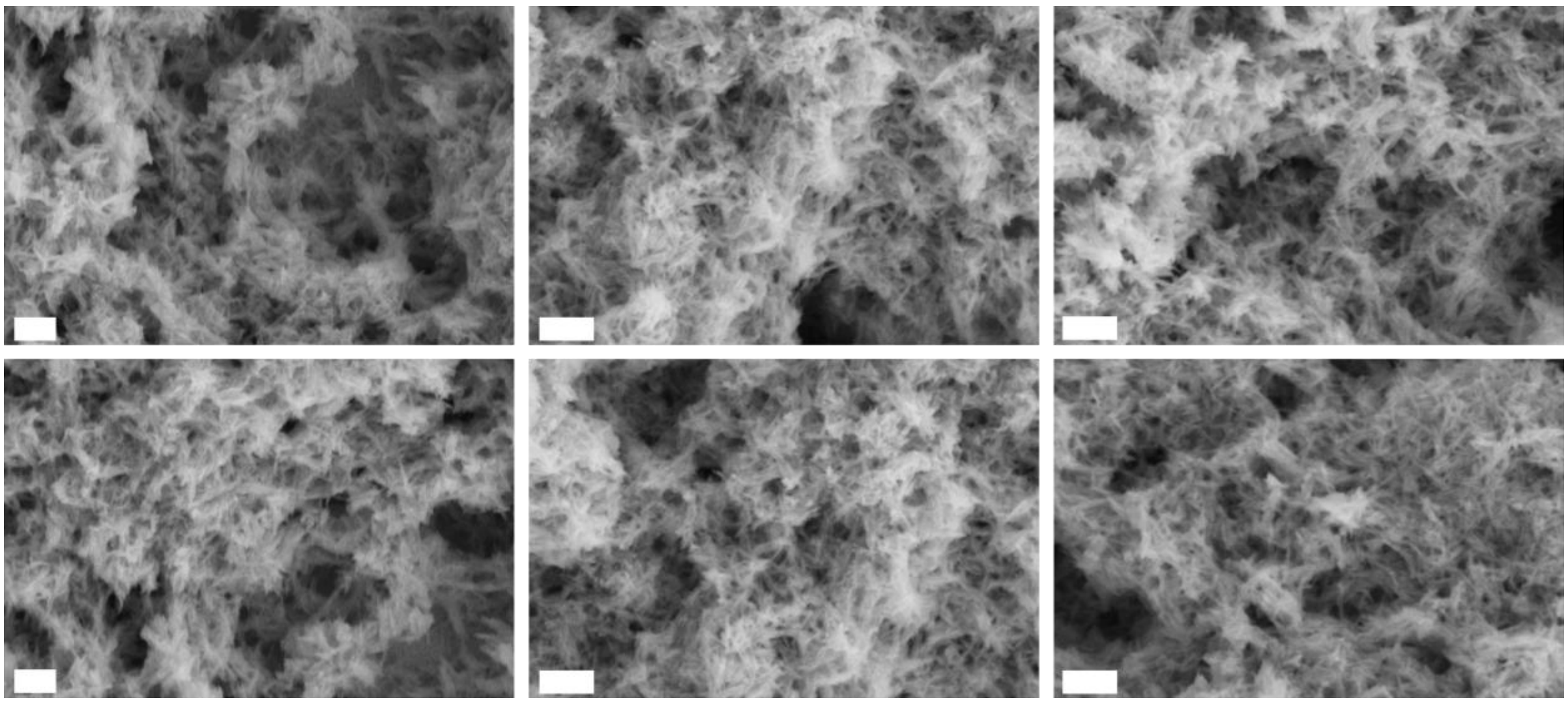

**Figure S7**. SEM images of the Te NPs taken with a secondary electron detector with an accelerating voltage of 5 kV and a working distance of 9.1 mm. The scale in all images is 200 nm.

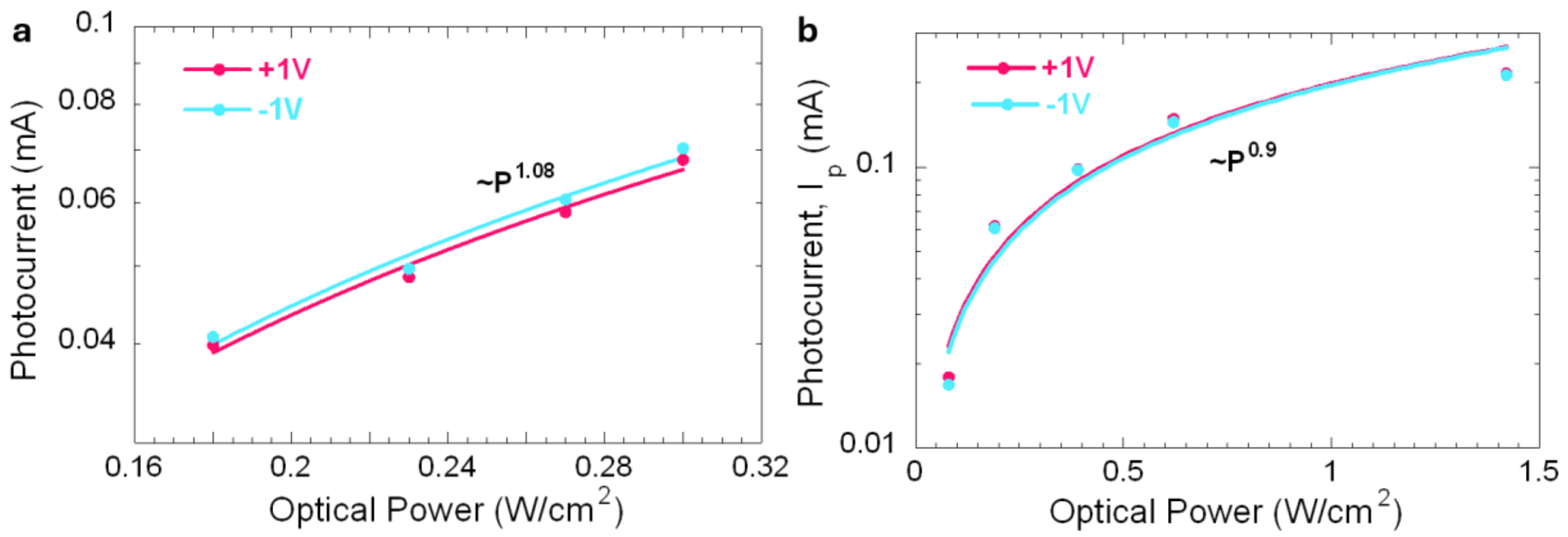

**Figure S8**. Photocurrent variation with optical power for illumination with light source of wavelengths **(a)** 405 and **(b)** 532 nm.

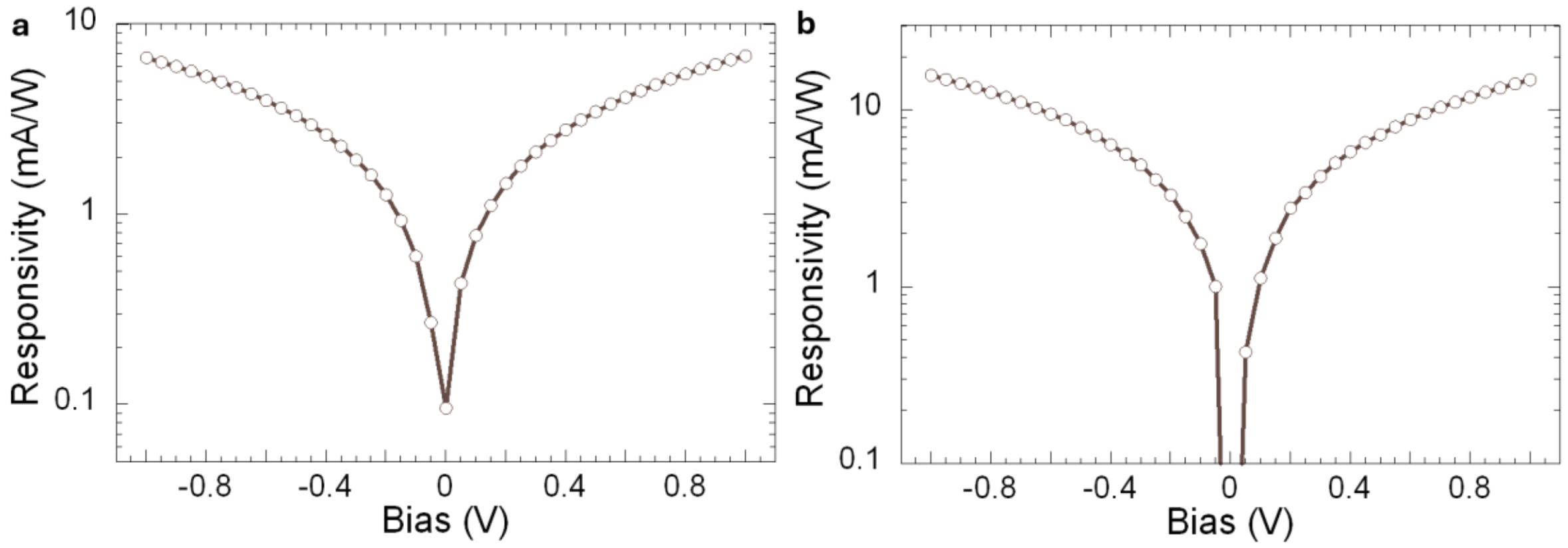

**Figure S9**. Bias dependence of responsivity for 1.42 W/cm$^2$ of 532 nm and 0.2 W/cm$^2$ of 405 nm source.

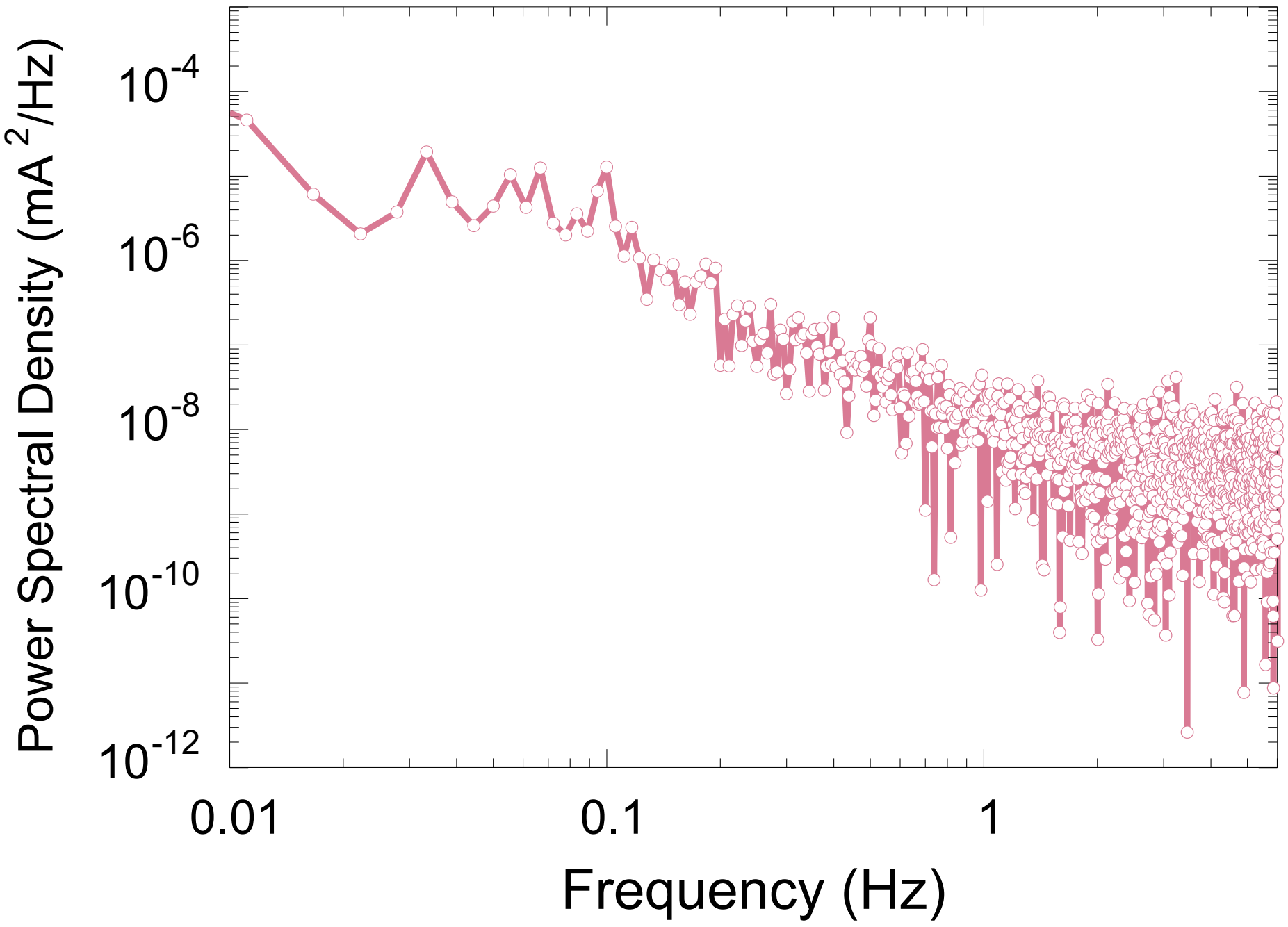

**Figure S10**. Power spectral density of the noise plotted as a function of noise frequency.

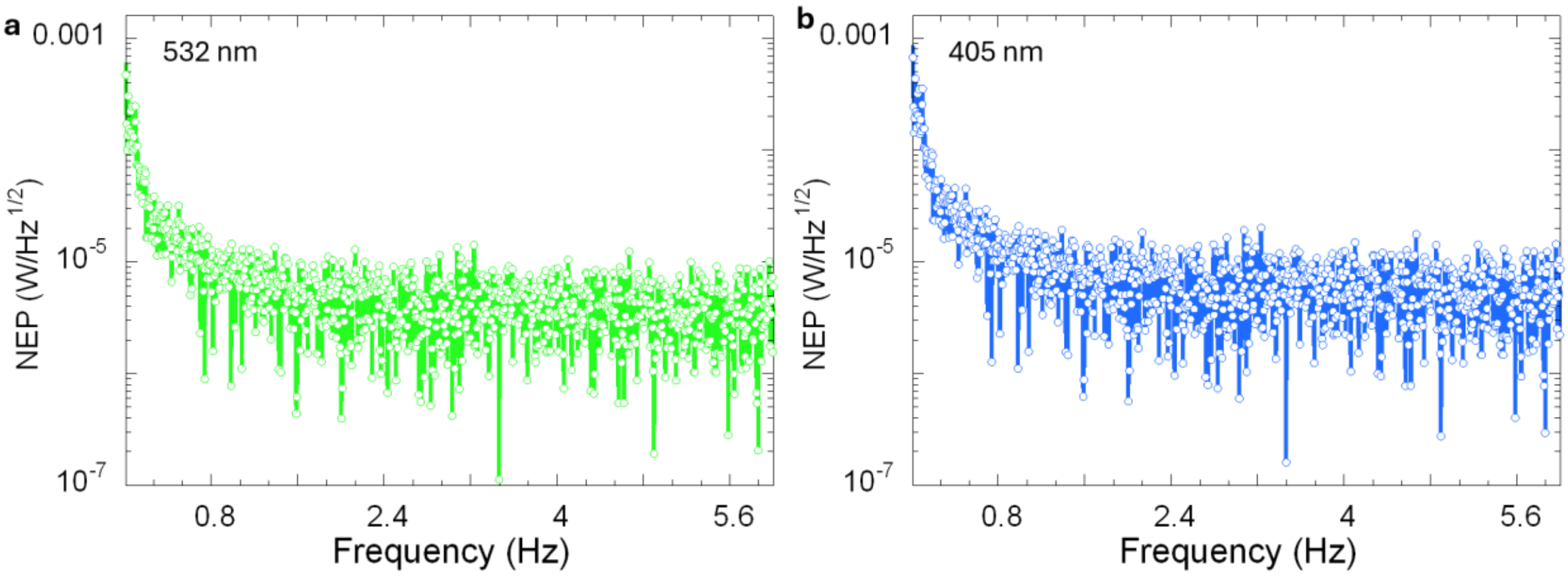

**Figure S11**. Noise equivalent power calculated for the highest responsivity obtained for **(a)** 532 nm (1.42 W/cm$^2$) and **(b)** 405 nm (0.2 W/cm$^2$) illumination.

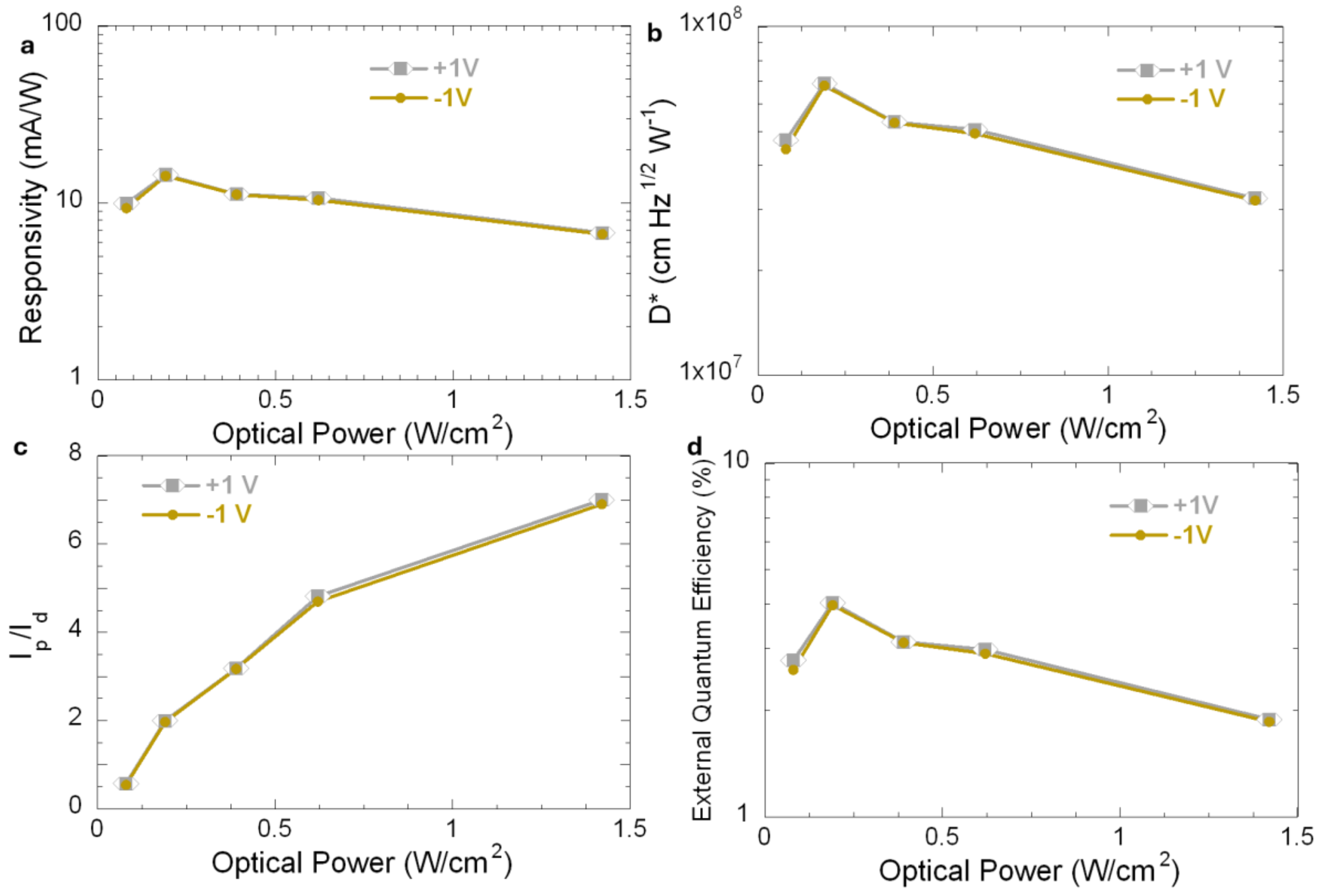

**Figure S12**. Device figures of merit for 532 nm illumination

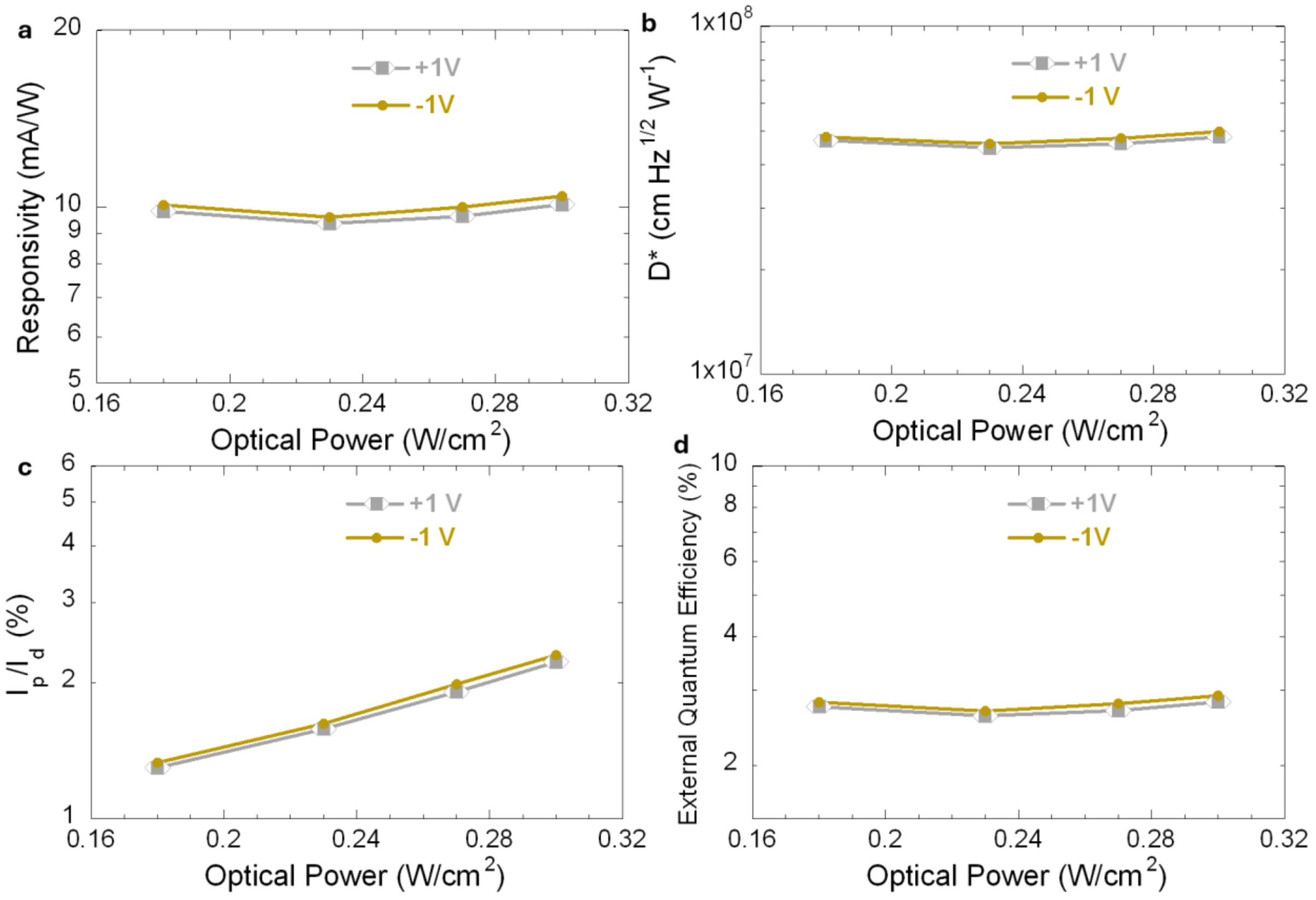

**Figure S13**. Device characteristics for 405 nm illumination

**Table S3: Tellurium-based Photodetection:**

| Material / Device architecture | Responsivity | EQE | Detectivity (Jones) | Rise / fall time |
|---|---|---|---|---|
| Present TNP ensemble | 8.5 mA/W | 2.4 % | $1.3 \times 10^{6}$ | 17.7 s / 40.2 s |
| Solution-processed, oxide-passivated Te nanoparticle films (2025)[3] | 1.24 mA/ W | 0.09 % | $8.9 \times 10^{7}$ | 90 ms / 103 ms |
| Hydrothermal Te Nanostructures (2026)[4] | | | | |
| Te nanoflakes (2020)[5] | 19.2 mA/W | 1.52% | NR (Not Reported) | Not directly reported; 37 MHz bandwidth |
| Te nanoflakes (2022)[6] | 51.85 A/W | 4148% | $1.88 \times 10^{10}$ | ~19 $\mu$s / 21 $\mu$s |
| CVD-grown Te nanowires (2021)[7] | 6650 A/W | NR | $1.23 \times 10^{12}$ | 31.7 $\mu$s / 25.5 $\mu$s |
| Epitaxial Te nanowires (2023)[8] | 327 A/W | NR | $6.08 \times 10^{7}$ | 22 $\mu$s / 23 $\mu$s |
| Hydrothermal Te flakes (2020)[9] | 19.2 mA/W | NR | NR | NR |
| $Se_{0.32}Te_{0.68}$ alloy thin film with optical cavity (2020)[10] | 1.5 A/W | NR | $6.5 \times 10^{10}$ | NR |
| Te/$MoS_2$ van der Waals heterojunction (2021)[11] | 4.0 A/W | NR | NR | NR |
| Metal–Te–metal ultrabroadband detector (2022)[12] | 9.38 A/W | NR | NR at 1550 nm | ~70 $\mu$s |
| Asymmetrically contacted Te nanobelt (2024)[13] | 26.1 A/W | NR | $3.24 \times 10^{9}$ | ~720 $\mu$s |
| Te/Si self-powered heterojunction (2024)[14] | 66 mA/W | 0.205% | $2.08 \times 10^{8}$ | NR |
| Te/$MoS_2$ vertically stacked heterostructure (2024)[15] | ~10 A/W under IR illumination | NR | NR at 1550 nm | NR |
| Te/$Bi_2Se_3$ van der Waals heterojunction (2025) [16] | 880 mA/W | NR | $1.85 \times 10^{10}$ | 0.12 / 0.156 ms |
| $Te_{0.65}Se_{0.35}$/$Al_2O_3$/GaAs heterojunction (2025)[17] | NR | NR | $1.8 \times 10^{9}$ | NR |
| Te/Si mixed-dimensional heterojunction (2025)[18] | $6.32 \times 10^{-4}$ mA/W | NR | NR | 2.4 / 2.4 ms |
| Graphene-electrode Te nanotubes (2026)[19] | 11 mA/W | NR | Not numerically specified | 0.397 / 0.344 ms |
| Te nanoribbon phototransistor (2025)[20] | ~$1.3 \times 10^{4}$ A/W | NR | ~$10^{15}$ | NR |
| Single Te nanowire using bulk photovoltaic effect (2026)[21] | 5.9 mA/W | NR | $1.85 \times 10^{11}$ | NR |
| Self-rolled Te photodetector (2024)[22] | ~81.6 V/W | NR | NR at 1550 nm | NR |
| Te/$PdSe_2$ van der Waals heterojunction (2024)[23] | 617 mA/W | NR | $5.27 \times 10^{10}$ | ~4.5 $\mu$s (response) |
| Te/$Bi_2O_2Se$ heterojunction (2026)[24] | ~0.89 A/W | NR | NR | ~29 / 41 $\mu$s |
| $Se_{0.3}Te_{0.7}$ photoconductor with $TeO_x$ passivation (2026)[25] | 57.7 mA/W | NR | NR | NR |
| Te-doped black-silicon photodiode (2020)[26] | 15 mA/W | NR | NR | NR |
| Te nanoparticle/PbS nanorod hybrid (2023)[27] | 15 mA/W | NR | NR | NR |

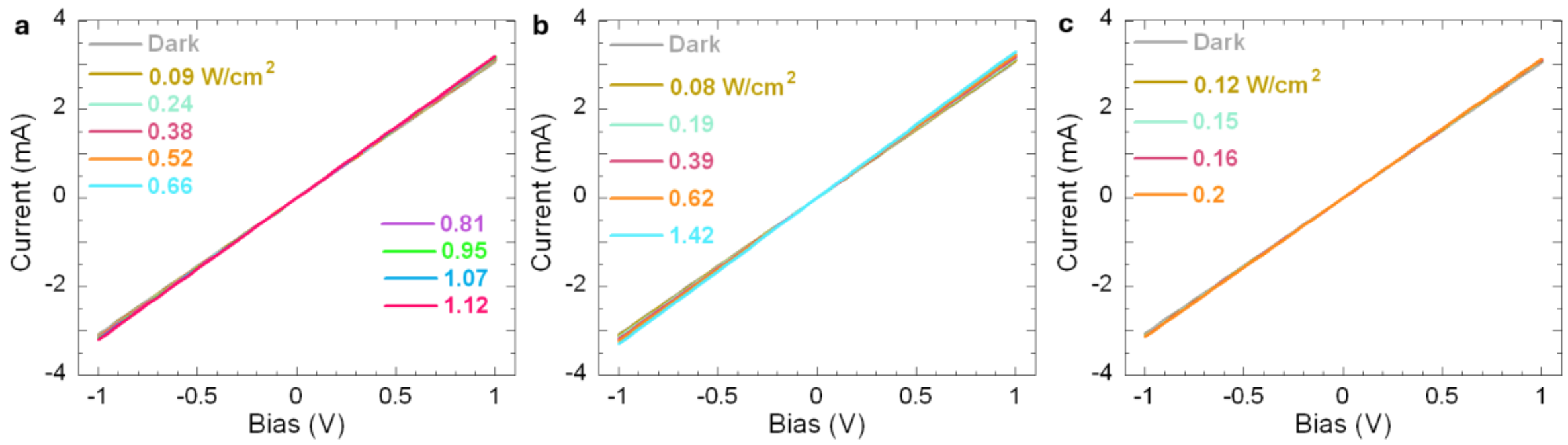

**Figure S14.** Current–voltage (I–V) characteristics of the device under dark and illuminated conditions at **(a)** 1550 nm, **(b)** 532 nm, and **(c)** 405 nm. The corresponding illumination power densities are indicated in the panels.

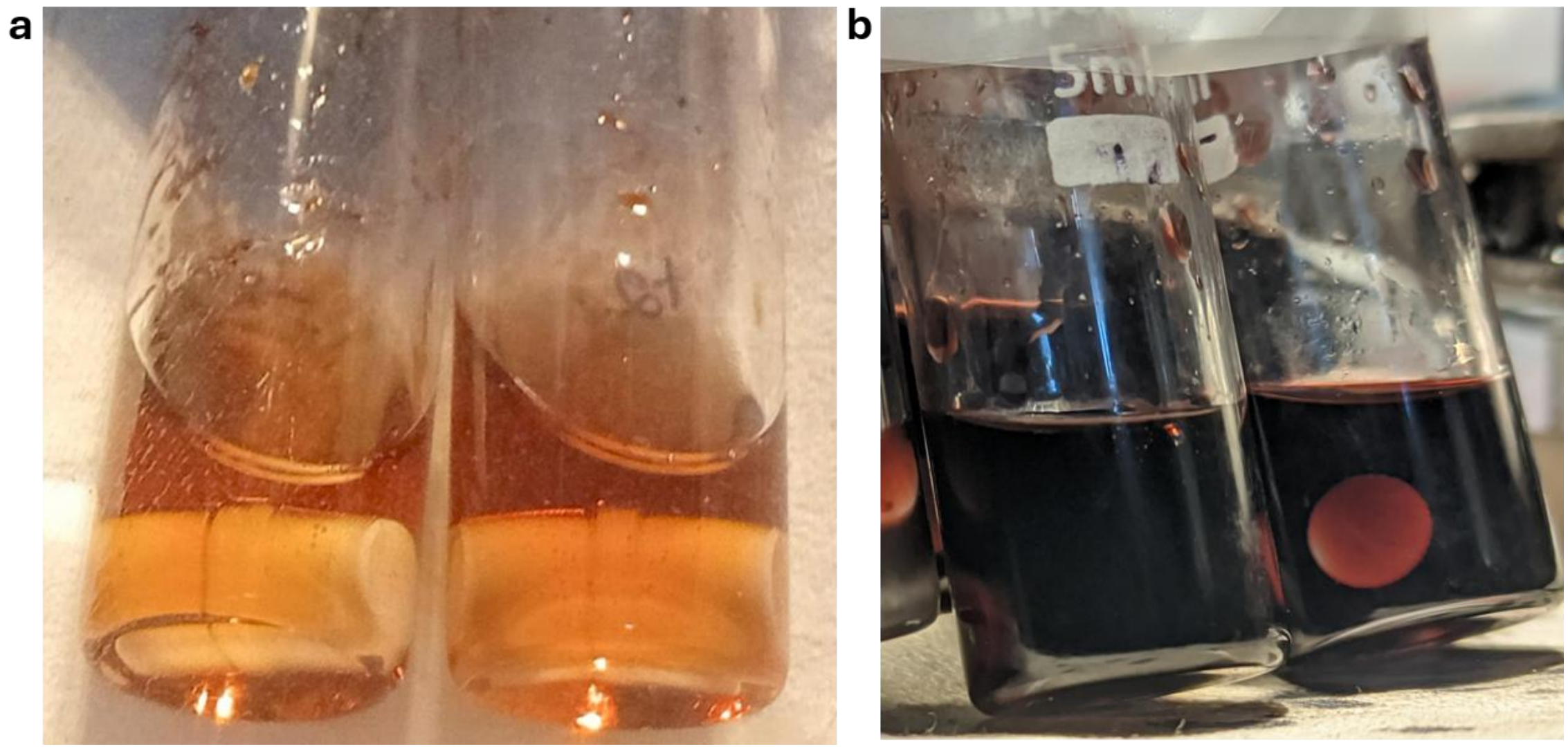

**Figure S15**. Photograph of the reaction mixture before and after air (oxygen )exposure for 2 hours.